\documentclass[preprint]{elsarticle}
\pdfoutput=1
\usepackage{amsmath}
\usepackage{latexsym}
\usepackage{graphicx}
\usepackage{listings}

\def\Sngp{\mathcal{S}^{(0)}}
\def\Snm{\mathcal{S}^{(l-1)}}

\def\Snp1{\mathcal{S}^{(l+1)}}
\def\Sn{\mathcal{S}^{(l)}}
\def\Sl{\mathcal{S}^{(1)}}
\def\Sq{\mathcal{S}^{(2)}}

\def\dS{\mathrm{d}\,{\bf S}}
\def\dV{\mathrm{d} \,V}
\def\dt{\mathrm{d} t}
\def\dx{\mathrm{d} x}
\def\bx{{\bf x}}

\def\bxp{{\bf x}_p}

\def\up{{\bf v}_p}
\def\B{{\bf B}}
\def\E{{\bf E}}

\def\bJ{{\bf J}}

\title{On Energy and Momentum Conservation in Particle-in-Cell Plasma Simulation}
\author{J. U. Brackbill }
\ead{jerrybrackbill@comcast.net}
\address{
Portland OR 97214 \\
{\it jerrybrackbill@comcast.net}}

\begin{document}
\begin{abstract}
Particle-in-cell (PIC) plasma simulations are a productive and valued tool for the study of nonlinear plasma phenomena, yet there are basic questions about the simulation methods themselves that remain unanswered.  Here we study one such question: energy and momentum conservation by PIC.   We employ both analysis and simulations of one-dimensional, electrostatic plasmas to understand why PIC simulations are either energy or momentum conserving but not both,  what the role of numerical stability is in non-conservation,  and how do errors in conservation scale with the numerical parameters.  Conserving both momentum and energy make it possible to model problems such as Jeans' -type equilibria.  Avoiding  numerical instability is useful, but so is being able to identify when its effect on the results may be important. Designing simulations to achieve the best possible accuracy with the least expenditure of effort requires results on the scaling of error with the numerical parameters..  Our results identify the central role of Gauss' law in conservation of both momentum and energy, and the significant differences in numerical stability and error scaling between energy-conserving and momentum-conserving simulations.
 \end{abstract}

\maketitle


\section{ Introduction}

Plasma simulations using particle-in-cell (PIC) methods  \cite{dawson1962, birdsall1985, hockney1988,grigoryev2002,dawson1995} have proven their value many times over.  They have helped several generations of plasma physicists understand fundamental phenomena, among them collisionless shocks \cite{forslund1970, forslund1984, lembege1989, lembege2004}, laser-plasma interactions \cite{,estabrook1983, forslund1975,  forslund1982, wilks1992}, and magnetic reconnection  \cite{,pritchett2001, ricci2004,joshi1984, mangles2004} in both the laboratory and in space plasmas.

Development of PIC simulation methods extends their range , improves their accuracy,  and adapts them to specialized applications.   There are now many different variants of PIC, including charge-conserving \cite{villasenor1992,chen2011} and implicit PIC simulation algorithms \cite{denavit1981,mason1981, brackbill1982,langdon1983,chen2011,markidis2011}. PIC simulations exist in many specialized forms,  but here we consider their common core in its simplest form to understand errors in momentum and energy conservation .   

PIC methods were developed originally to model fluid flows in two dimensions\cite{harlow1959,harlow1964}.  PIC plasma simulations and PIC fluid algorithms are similar in many ways.  However, where PIC fluid models conserve both momentum and energy, PIC plasma simulations conserve either momentum or energy but not both. The errors are not necessarily large, especially if numerical instability can be avoided, but there are problems where non-conservation of momentum or energy can be a problem.  For example, kinetic equilibria based on Jeans' theorem  \cite{schindler1973} require accurate conservation of both momentum and energy. 

The energy and momentum conservation properties of PIC simulations are discussed in \cite{birdsall1985, hockney1988}, and a striking example of momentum non-conservation is given in \cite{langdon1973}.
In \cite{hockney1988}[p252],  errors in energy conservation in momentum-conserving schemes and errors in momentum conservation in energy-conserving schemes are attributed to the same cause, 'the nonconservative part of the force due to undersampling'.  In a more recent paper \cite{markidis2011}[p7041], errors in momentum conservation are attributed to 'spurious self-forces arising from the non-smoothness of the current deposition to the grid.'  Even an energy and charge-conserving method does not enforce exact momentum conservation \cite{chen2011}.
 
 To understand why there are errors in conservation in PIC plasma simulations, we consider simulations in their simplest form:  one-dimensional, electrostatic simulations with periodic boundary conditions.   This problem is simple enough to analyze, and modest enough to explore error scaling with numerical parameters  yet complex enough to be relevant. 
We review the equations describing an electrostatic plasma, and identify what  a numerical model must get right if the energy and momentum are to be conserved.  We compare results for
LEWIS, an energy-conserving method using point particles and a quadratic variation of the potential; CIC, a  standard momentum-conserving, cloud-in-cell (CIC) algorithm; and CELESTE, a contemporary version of Lewis' energy conserving method \cite{vu1992, sokolov2013, markidis2011, chen2011, chacon2013} that uses particle clouds as in CIC and a linear potential variation.  
With all three methods, we compare the stability of simulations with dispersion theory \cite{langdon1970,lindman1970}, and examine the scaling of momentum and energy errors with the time step, grid spacing, and number of particles.
 
Our principal results are:
\begin{itemize}
\item Momentum is conserved to round-off when Gauss' law is satisfied at grid points where the electric field is stored.  
\item Re-centering the electric field by averaging is essential to momentum conservation in CIC.  Averaging has consequences:
 \begin{itemize}
 \item There is introduced an energy error that is independent of the time step,  $\Delta t$.  
 \item Cold beams are unstable at all speeds  \cite{birdsall1980}.    The instability is not consistent with the dispersion theory for the finite grid instability  \cite{langdon1970,lindman1970}.
 \end{itemize}
\item The accuracy of  energy-conserving methods is limited by the accuracy of the solutions of the particle equations of motion and the interpolation order. 
 \end{itemize}

 In outline, the paper is organized as follows:  In Model, the equations for an electrostatic plasma and their conservation properties are discussed.  In Algorithms, three numerical algorithms, LEWIS, CIC and CELESTE are developed, and their momentum conservation properties are discussed.  In Analysis, the numerical stability and errors in energy conservation are reviewed.  In Numerical Test Problems, the results of simulations for a cold electron beam, a warm stationary plasma, and a drifting warm plasma 
 are used to develop scaling.  In Conclusions, we summarize the results.

\section{The Equations for an Electrostatic Plasma in One Dimension and Their Properties}
\label{Model}

We review the equations for an electrostatic plasma to remind ourselves that conservation of charge, momentum and energy are fundamental properties of their solutions. 

The equations for  an electrostatic plasma describe the motion of charged particles in the electric field they induce. Each particle $p$ has position, velocity,  and charge, $\bxp$,  $\up$,  and  $q_p$, and the electric field is derivable from a scalar potential, $\E=-\nabla \phi$.

The charge density is,
\begin{equation}
Q(\bx)=\sum_p q_p \delta(\bx-\bxp) ,
\label{Qdef}
\end{equation}
 where $\delta(x)$ is a Dirac delta function, and $\delta(\bx)=\delta(x) \delta(y) \delta(z)$.  We assume the total charge within a volume changes only as a result of particles entering or leaving  the volume. 

The equations of motion are derived in the classical way from Hamilton's principle for charged particles in an electrostatic field \cite{goldstein1959} [pp364-370], for which the Lagrangian is,
\begin{equation}
\mathcal{L}= \mathcal{F} + \mathcal{K}-  \Phi.
\label{lagrangian}
\end{equation} 
The Lagrangian comprises the field, 
\begin{subequations}
\begin{equation}
 \mathcal{F} = \frac{1}{8 \pi} \int_D \mathrm{d} ^3 x \,  \nabla \phi^2,
 \label{fieldNRG}
 \end{equation}
particle kinetic,
\begin{equation}
 \mathcal{K}= \sum_p \frac{1}{2} m_p v_p^2,
 \label{kineticNRG}
 \end{equation}
and particle potential energies,
\begin{equation}
\Phi= \int_D \mathrm{d}^3x \,Q(\bx) \,\phi(\bx).
\label{potentialNRG}
\end{equation}
\end{subequations}
The potential energy links the field and particle energies.  

Variation of $\mathcal{L}$ with respect to the $\phi$ yields Poisson's equation and Gauss' law, 
\begin{equation}
4 \pi Q(\bx) = -\nabla^2 \phi = \nabla \cdot \E.
\label{modelPoisson}
\end{equation}

Variation with respect to the particle position yields the particle equations of motion,
\begin{eqnarray}
m_p \frac{\mathrm{d} \up}{\dt}& = & q_p \E (\bxp)  \nonumber \\
\frac{\mathrm{d} \bxp }{\dt} & = &  \up.
\label{Lorentz}
\end{eqnarray}


\subsection{momentum}
The total particle momentum,
$$ {\bf{P }}  \equiv \int_D \sum_p m_p{ \up} \delta(\bx-\bxp) \dV ,$$ 
changes only as a result of forces acting on the boundary $\partial D$.
The time derivative of the total momentum, 
\begin{equation}
 \frac{\partial \bf{P}}{\partial t}=\int _D Q \, {\E} \mathrm{d} V, 
 \label{nonconservative}
 \end{equation}
is a conservation law because Gauss' law, Eq. \ref{modelPoisson}, is satisfied. With Gauss' law, the RHS is,
$$\left( \nabla \cdot \E \right)  \E = Q\E, $$
which for an electrostatic plasma where $\E \times (\nabla \times \E) =0$ is the Maxwell stress \cite{landau1975}, 
\begin{equation}
 \nabla \cdot \left( \E \E -  \mathbf{I} \frac{\E \cdot \E}{2}\right)=\left ( \nabla \cdot \E \right) \E.
 \label{EMaxwell}
 \end{equation}
($\mathbf{I}$ is the unit tensor.) It is true even when $\nabla \times \E \neq 0$ \cite{landau1975}. 

Combining Eqs. \ref{nonconservative} and \ref{EMaxwell} shows the system momentum changes only as the result of boundary contributions,
\begin{equation}
 \frac{\partial \bf{P}}{\partial t}= \frac{1}{4 \pi} \int_{\partial D} \dS  \cdot \left( \E \E -  \mathbf{I} \frac{\E \cdot \E}{2}\right) . 
 \label{conserveP} 
 \end{equation}
If the boundaries are periodic or $\hat{\bf n} \cdot \E=0$,
the total particle momentum is a constant of the motion.

The Maxwell stress for a vacuum electromagnetic field is derived in \cite{landau1975}[pp86-87].   For a magnetized plasma, Eq. \ref{conserveP} becomes,
$$ \frac{\partial \bf{P'}}{\partial t} = \frac{1}{4 \pi} \int_{\partial D} \dS  \cdot \left( \E \E + \B \B -  \mathbf{I} \frac{\E \cdot \E+\B \cdot \B}{2}\right), $$
where ${\bf P}'={\bf P}+\int_D \mathrm{d} V (\E \times \B)/4\pi c .$  In this derivation, the solenoidality condition for the magnetic field, $ \nabla \cdot \B=0,$ and Gauss's law have similar roles  \cite{brackbill1980}.

\subsection{energy}
\label{ModelNRG}
Energy is conserved.  Where $\mathcal{E} =  \mathcal{K}+ \mathcal{F}$, changes in $\mathcal{E}$ are given by,
$$\frac{\partial}{\partial t} ( \mathcal{K} + \mathcal{F})=\int_D  \E  \cdot \left( \bJ+\frac{1}{4 \pi} \frac{ \partial \E} {\partial t} \right)  \dV, $$
with current density,
$$\bJ=\sum_p q_p \up \delta(\bx-\bxp).  $$
Substitute $-\nabla \phi$ for $E$, and and integrate by parts to derive, 
\begin{equation}
-\int_D  \phi  \nabla \cdot \left( \bJ+\frac{1}{4 \pi} \frac{ \partial \E} {\partial t} \right)  \dV=-\int_{\partial D}\dS \cdot \left( \bJ+\frac{1}{4 \pi} \frac{ \partial \E} {\partial t} \right) \phi .
\label{NRG}
\end{equation}
In this form the RHS is the boundary contribution. 

Consider a periodic domain, for which the RHS  vanishes. Unless $\phi$ is equal to zero everywhere, energy is conserved because  charge  is conserved ,
\begin{equation}
0=\nabla \cdot \bJ + \partial Q/\partial t,
\label{continuity}
\end{equation}
and charge conservation enters because the divergence term  on the LHS of Eq. \ref{NRG}
is the time derivative of 
Gauss' law, Eq. \ref{modelPoisson}, which is always satisfied.

To point out the obvious, the negative of this proposition is also true.  If the charge continuity equation were not satisfied, energy would not be conserved.

\section{Electrostatic Particle-in-Cell (PIC) Algorithms in One Dimension}
\label{Algorithms}

The original particle-in-cell (PIC) method models hydrodynamic flow  by the motion of  Lagrangian particles through a stationary grid in a single-valued, continuous flow velocity \cite{harlow1959, harlow1963, harlow1964, amsden1966} .  The particle motion is computed by area-weighting the velocities at nodes of the grid \cite{harlow1959}, and solutions conserve both momentum and energy. Accuracy is compromised by a 'ringing instability' that causes diffusion, especially in stagnation regions \cite{harlow1963, brackbill1988}.  Newer versions of PIC reduce or eliminate the instability with smoother interpolation \cite{brackbill1986b}.

PIC plasma simulation models collision-less plasmas, for which the electric and magnetic fields are continuous but velocities are not \cite{birdsall1985}. Plasma motion is modeled by the motion of computational particles through a stationary grid on which continuous electric and magnetic fields are computed.  The particle motion is computed by area-weighting the fields at nodes of the grid.  PIC simulations presently conserve either momentum or energy.  Simulation has solved many fundamental plasma problems, but its accuracy is compromised by a finite-grid-instability (FGI) that causes diffusion and heating \cite{langdon1970, lindman1970}. 

Here we study the properties of 3 PIC simulation methods; the cloud-in-cell method (CIC) \cite{birdsall1985}, Lewis' point-particle energy-conserving method (LEWIS) ]cite{Lewis1970}, and a cloud-in-cell method that is also energy-conserving (CELESTE) \cite{vu1992}.   We limit our study to an electrostatic plasma  in one dimension on a periodic domain $x\in[0,L]$.  The domain is divided into $N$ cells with width $\Delta x$.  The values of the electric field, $E_v$, are calculated at cell vertices,  $x_v: v \in [1,N+1]$, from a potential, $\phi_c$, at  cell centers $x_c=1/2(x_{v+1}+x_v)$.    
 All the algorithms execute the same sequence of steps to solve initial value problems.  \begin{itemize}
 \item   A charge density is computed at grid points. (Sec. \ref{chargeassignment} )
\item   A finite-difference approximation to Poisson's equation is solved to compute the electric field.   ( Sec. \ref{Poisson})
\item   The electric field is interpolated from the grid to the particles. (Sec \ref{forceinterpolation})
\item   Particle equations of motion are solved using leapfrog time-stepping.  (Sec. \ref{particlemotion})
\end{itemize}
 The three methods transform the equations of motion for an electrostatic plasma into a PIC formulation differently, but all use the special properties of the b-spline,  which are described in many references, among them \cite{deboor1978,haugbolle2013}. In Appendix we list relevant properties b-splines for our comparison of  PIC simulation methods.

\subsection{Charge assignment}
\label{chargeassignment}
 
The cloud-in-cell (CIC) method \cite{birdsall1985} substitutes '{\bf clouds}' for the point particles in Section \ref{Model}.  The clouds move with unchanging shape and size at their center-of-mass velocity. Mathematically, a nearest-grid-point (NGP) b-spline replaces Dirac delta functions, and the charge density, Eq. \ref{Qdef}, is replaced by,
\begin{equation}
Q(x)=\sum_p q_p \Sngp(x-x_p)
\label{QCIC}
\end{equation}  
The CIC charge is assigned in proportion to the overlap of cloud and cell, 
\begin{equation} 
Q_c^{(1)}=\sum_p q_p \int \dx \Sngp(x-x_p) \Sngp(x-x_c). 
\label{CICQ}
\end{equation}
We recognize this as the convolution that defines $\Sl$,  Eq. \ref{convolution},  and the assignment in  Eq. \ref{CICQ} uses a linear b-spline with $l=1$.

Lewis' PIC method \cite{lewis1970}, which we will call LEWIS, follows the motion of {\bf point particles} in a  potential whose values at  a finite number of grid points are the unknowns, and $\phi(x)$ is given an expansion in b-splines,
\begin{equation}
\phi(x)=\sum_c \phi_c \Sn(x-x_c),
\label{phiLewis}
\end{equation}
where $\phi_c$ is the value of the potential at cell centers.  With the charge density for point particles, Eq. \ref{Qdef}, the particle potential energy, Eq. \ref{potentialNRG} is then given by,
\begin{equation}
\Phi=\int \dx \sum_p \delta(x-x_p) \sum_c \phi_c \Sn(x-x_c),
\end{equation}
or 
\begin{equation}
\Phi=\sum_c Q_c^{(l)} \phi_c \Delta x,
\label{PhiLewis}
\end{equation}
where $Q_c^{(l)}$ is the LEWIS charge density in cell $c$, 
\begin{equation}
Q^{(l)}_c  =  \sum_p q_p \Sn(x_c-x_p) 
\label{Step1Lewis}
\end{equation}

CELESTE follows the motion of {\bf clouds} in a potential given by Eq. \ref{phiLewis} with $l=1$.
The particle potential energy, Eq. \ref{potentialNRG}, becomes,
\begin{equation}
\Phi=\int \dx \sum_p \Sngp(x-x_p) \sum_c \phi_c \Sl(x-x_c).
\end{equation}
The convolution results in a particle potential energy,
\begin{equation}
\Phi=\sum_c Q_c^{(2)} \phi_c,
\label{PhiCELESTE}
\end{equation}
where $Q_c^{(2)}$, the CELESTE charge density in cell $c$, is given by Eq. \ref{Step1Lewis} with $l=2$.

\subsection{Poisson's equation}
\label{Poisson}

In CIC, the values of the potential at cell centers are obtained by solving a finite difference approximation to Poisson's equation,
\begin{equation}
4 \pi Q_c^{(1)} =-\frac{\phi_{c+1}-2 \phi_c + \phi_{c-1}}{\Delta x^2}.
\label{PoissonCIC}
\end{equation}
With a vertex electric field defined by,
\begin{equation}
E_v \equiv -\frac{\phi_c-\phi_{c-1)}}{\Delta x},
\label{E_v}
\end{equation}
Gauss' law for CIC is,
\begin{equation}
4 \pi Q_c^{(1)} =\frac{E_{v+1}-E_v}{\Delta x}.
\label{GaussCIC}
\end{equation}

In LEWIS, Hamilton's principle \cite{goldstein1959} is used to derive an equation for the potential.  The field energy, Eq. \ref{fieldNRG}, becomes,
\begin{equation}
 \mathcal{F}= \frac{1}{8 \pi} \sum_{c,c'} \left( \frac{\phi_{c} - \phi_{c-1}}{\Delta x} \right) \left(\frac{\phi_{c'}-\phi_{c'-1}}{\Delta x} \right) \Sn_{cc'}  \Delta x, 
 \label{LewisF}
 \end{equation}
The 'mass matrix' in Eq. \ref{LewisF} is,
\begin{equation}
\Sn_{cc'} \Delta x= \int_0^L \dx \Snm (x-x_c) \Snm(x-x_{c'}). 
\label{Lewismass}
\end{equation}
The Euler-Lagrange equation for the potential is obtained by varying the particle potential energy, Eq. \ref{PhiLewis} and field energy, Eq. \ref{LewisF}, with respect to $\phi_c$.  The result is the LEWIS version of Poisson's equation,
\begin{equation}
4 \pi Q^{(l)}_c = -\sum_{c'} \frac{\phi_{c'+1} -2 \phi_{c'} + \phi_{c'-1}}{\Delta x^2} \Sn_{cc'}.
\end{equation}
A 'sharpened' density, given by the solution of,
\begin{equation}
\sum_{c'} \widehat{Q_c^{(l)}} \Sn_{cc'} =Q_c^{l)},
\end{equation}
puts Poisson's equation in a form that can be solved by standard methods,
\begin{equation}
4 \pi \widehat{Q^{(l)}_c} = -\frac{\phi_{c+1} -2 \phi_{c} + \phi_{c-1}}{\Delta x^2}.
\label{Step2Lewis}
\end{equation}
With $E_v$ defined by Eq. \ref{E_v} Gauss' law is,
\begin{equation}
4 \pi  \widehat{Q^{(l)}_c} =\frac{E_{v+1}-E_v}{\Delta x}.
\label{GaussLewis}
\end{equation}

For $l=1$, the mass-matrix reduces to a Kronecker $\delta$,  $\mathcal{S}^{(1)}_{cc'}=\delta_{c,c'}$, $\widehat{Q_c^{(1)}}=Q_c^{(1)}$, and Poisson's equation for LEWIS (and CIC) are given by Eq. \ref{PoissonCIC}.  For $l=2$, $\mathcal{S}^{(2)}_{cc'}=\mathcal{S}^{(3)}(x_c-x_{c'}) $ \cite{langdon1973}, and the sharpened density is given by solving,
$$1/6 \widehat{Q_{c+1}} + 2/3\widehat{ Q_c}+ 1/6 \widehat{Q_{c-1}}=Q_c^{(2)}.$$

For CELESTE, the field energy is given by Eq. \ref{LewisF} with $l=1$ and the particle potential energy by Eq. \ref{PhiCELESTE}.  Variation with respect to $\phi_c$ yields Poisson's equation with charge density $Q_c^{(2)}$ and no mass matrix,
\begin{equation}
4 \pi Q_c^{(2)}=-\frac{\phi_{c+1}-2 \phi_c + \phi_{c-1}}{\Delta x^2}.
\end{equation}
Gauss' law for CELESTE is
\begin{equation}
4 \pi Q_c^{(2)} =\frac{E_{v+1}-E_v}{\Delta x}.
\label{GaussCELESTE}
\end{equation}

\subsection{Electric field interpolation}
\label{forceinterpolation}

CIC assigns the electric field to the particle positions using area-weighting, 
\begin{equation}
E(x_p)= \int \dx \sum_c E_c \Sngp (x-x_c) \Sngp (x-x_p),
\label{Step3CIC}
\end{equation}
the convolution of a piecewise constant electric field and a cloud.  Thus, the CIC electric field is an expansion in a linear b-splines,
  \begin{equation}
E(x_p)=\sum_c E_c \Sl(x_p-x_c).
\label{Step3CIC}
\end{equation}
The cell-centered field, $E_c$,  is computed by averaging $E_v$ ,
\begin{equation}
E_c=1/2 \left( E_{v+1}+E_v \right).
\label{E_c}
\end{equation}

For LEWIS (and CELESTE) recall that $\mathcal{S}$  is differentiable, 
the electric field is computed analytically, $E=-\partial \phi/\partial x$, and $x_c = x_{v}+\Delta x/2$ to derive,
\begin{equation}
E(x)=- \frac{\partial \phi}{\partial x} =\sum_v E_v \mathcal{S}^{(l-1)}(x-x_v),
\label{Step3Lewis}
\end{equation}
With $l=1$, the electric field is piecewise constant.  That means that with LEWIS' point particles, the electric field is assigned to particles using a nearest grid point function, 
\begin{equation}
E(x_p)=\sum_v E_v \Sngp(x_p-x_v).
\label{ENGP}
\end{equation}
but with CELESTE as in CIC, the force on a particle is the convolution two nearest-grid-point functions and the electric field is an expansion in linear b-splines,
  \begin{equation}
E(x_p)=\sum_v E_v \Sl(x_p-x_v).
\label{Step3CELESTE}
\end{equation}
This is also the expansion for LEWIS with $l=2$.

Results for LEWIS with $l=1$, with the electric field given by Eq. \ref{ENGP} are described as noisy in \cite{okuda1972,langdon1973}.  Therefore we include only LEWIS with $l=2$ in the comparisons between CIC, CELESTE, and LEWIS that follow. 

\subsection{Particle equations of motion}
\label{particlemotion}

The variation of the particle kinetic and potential energies yields the particle equations of motion.  The particle kinetic energy, Eq. \ref{kineticNRG}, is,
\begin{equation}
\mathcal{K}= \sum_p \frac{1}{2} m_p v_p^2,
\label{LewisK}
\end{equation}
and the particle equations of motion with leapfrog differencing are,
\begin{subequations}
\begin{eqnarray}
\frac{v_p^{n+1/2}-v_p^{n-1/2}}{\Delta t}&  =  & q_p E(x_p^n),   \\
\frac{x_p^{n+1}-x_p^n}{\Delta t}& =& v_p^{n+1/2}. 
\label{Lewis}
\end{eqnarray}
\label{Step4Lewis}
\end{subequations}

The particle electric fields are interpolated using a linear b-spline for CIC, LEWIS and CELESTE.  However, the CIC electric field is given by Eq. \ref{E_c} and the CELESTE and LEWIS field by Eq. \ref{E_v}.

\subsection{momentum conservation}
\label{momentumconservation}
Following the steps in the discussion of momentum conservation, it becomes obvious that LEWIS and CELESTE are not momentum conservative.  The change in total particle momentum each time step is,
\begin{equation}
\Delta \mathcal{P}= \sum_p q_p E(x_p^n) \Delta t.
\label{LewisP}
\end{equation}
With the electric field given by Eq. \ref{Step3CELESTE} ,  the RHS of Eq. \ref{LewisP} becomes,
\begin{equation}
\sum_v Q_v^{(1)} E_v \Delta x \Delta t=\sum_p q_p E(x_p) \Delta t.
\label{RHSLewis}
\end{equation}
The charge density, $Q_v^{(1)}=\sum_p q_p \Sl(x_v-x_p)$, is not the same as the charge density  in Gauss' law, $\widehat{Q_c^{(2)}}$, Eq. \ref{GaussLewis}, or $Q^{(2)}$ in Eq. \ref{GaussCELESTE}, the RHS in Eq. \ref{RHSLewis} cannot be written in the conservation form given in Eq. \ref{conserveP}, and momentum is not conserved unless $Q_v^{(1)}=0$ for every $v$.

CIC conserves momentum.  With a cell-centered electric field, Eq. \ref{Step3CIC}, the RHS of Eq. \ref{LewisP} becomes,
\begin{equation}
 \sum_c Q_c^{(1)} E_c \Delta x \Delta t=\sum_p q_p E(x_p^n) \Delta t,
\end{equation}
the charge density is the same as in Gauss' law, Eq. \ref{GaussCIC}, and 
\begin{equation}
 \frac{1}{8 \pi} \sum_c (E_{v+1}^2-E_v^2) =\sum_c Q_c^{(1)} E_c
\end{equation}
is a collapsing sum that is zero for periodic boundary conditions, and a function only of the boundary conditions on the electric field  otherwise.

\section{Analysis}
\subsection{finite-grid-instability}
\label{aliasing}

The finite grid instability in PIC plasma simulations of a cold, drifting plasma \cite{langdon1970, lindman1970} and the ringing instability in PIC fluid modeling \cite{harlow1963} are both caused by aliasing and have similar properties \cite{brackbill1988}.  Because  there are many more particles than grid points, there is a degeneracy that causes numerical instability.  

The general form of the linear dispersion with alias contributions is \cite{birdsall1980},
\begin{equation}
0=1-\frac{\omega_p^2}{K^2(k)} \sum_q \frac{k_q \kappa(k_q) \mathcal{S}_k^{l+1}(k_q \Delta x/2)}{(\omega-k_q U_0)^2},
\label{FGIdisp}
\end{equation}
where 
$$ \mathcal{S}_k(k y)=\left(\frac{sin(ky)}{ky}\right),  $$
and  $-\pi /\Delta x \leq k \leq \pi/\Delta x$,  $k_g \equiv 2 \pi/\Delta x$, and $k_q=k-q k_g$, $q=0, \pm 1, \pm 2, \pm 3, ...$,   .  $K(k)$ is the Fourier transform of the finite-difference Laplace operator,  $\kappa(k_q)$ is the Fourier transform of the finite-difference gradient operator, and  $\mathcal{S}_k^{l+1}$ is the Fourier transform of $\Sn$ in $Q_c^{(l)}$. The solutions, in the form of complex $\omega$, are the roots of the dispersion relation.


For CIC, $K(k)=k^2 \mathcal{S}^2_k(k \Delta x/2)$, $\kappa(k_q)= k_q \mathcal{S}_k (k_q \Delta x)$, and charge is assigned with a linear b-spline, $l=1$.  Eq. \ref{FGIdisp} reduces to,
\begin{equation}
0=1-\omega_p^2 \sum_q \frac{\mathcal{S}_k (k_q \Delta x)}{(\omega-k_q U_0)^2}.
\label{CICFGI}
\end{equation}
For LEWIS  with $l=2$, $K(k)=k^2 \mathcal{S}^6_k(k \Delta x/2)$, $\kappa(k_q)= k_q \mathcal{S}_k (k_q \Delta x/2)$, and the charge is assigned with a quadratic b-spline.   Eq. \ref{FGIdisp} reduces to,
\begin{equation}
0=1-\frac{\omega_p^2}{\mathcal{S}^4_k (k \Delta x/2)}\sum_q \frac{\mathcal{S}^2_k (k_q \Delta x/2)}{(\omega-k_q U_0)^2}.
\label{LewisFGI}
\end{equation}
For CELESTE  $K(k)=k^2 \mathcal{S}^2_k(k \Delta x/2)$, $\kappa(k_q)= k_q \mathcal{S}_k (k_q \Delta x/2)$, and the charge is assigned with a quadratic b-spline. Eq. \ref{FGIdisp} reduces to,
\begin{equation}
0=1-\omega_p^2 \sum_q \frac{\mathcal{S}_k^2(k_q \Delta x/2)}{(\omega-k_q U_0)^2}.
\label{PICeFGI}
\end{equation}

The finite grid instability growth rate does not depend on the number of particles per cell or the time step.  

\begin{figure}
\begin{tabular}{c}
(a) \\
\includegraphics[width=90mm]{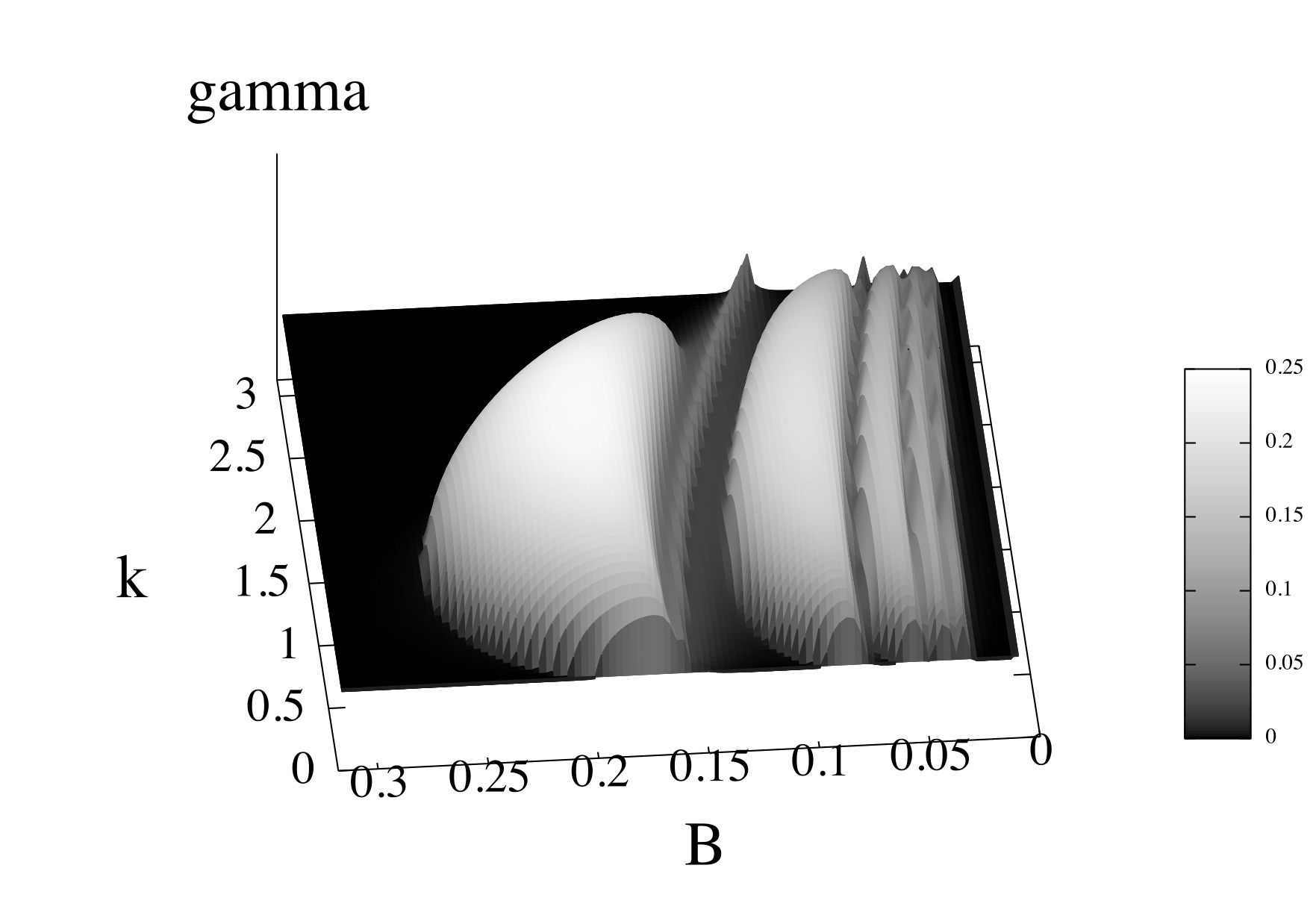} \\
(b) \\
\includegraphics[width=90mm]{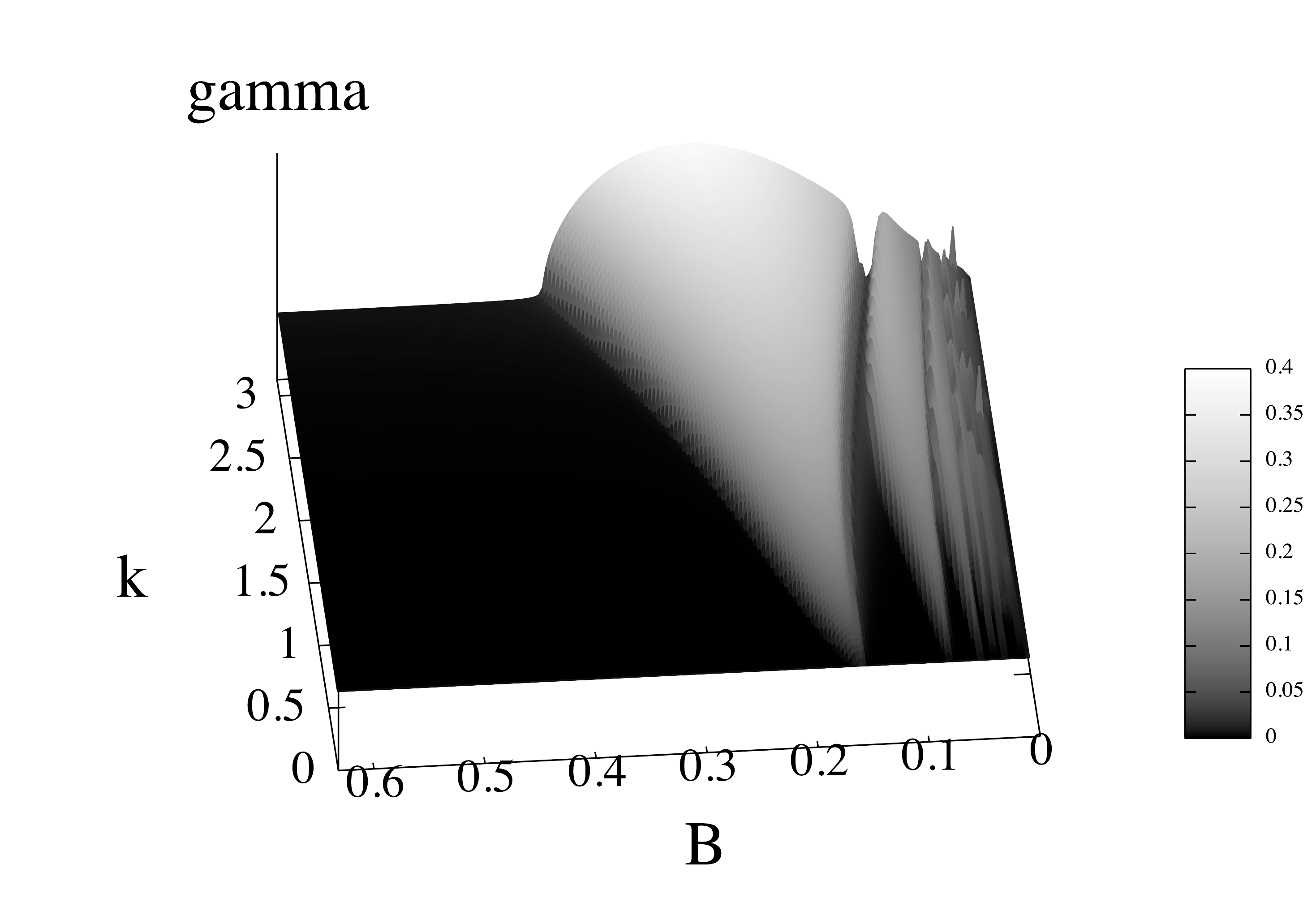} \\
(c) \\
\includegraphics[width=90mm]{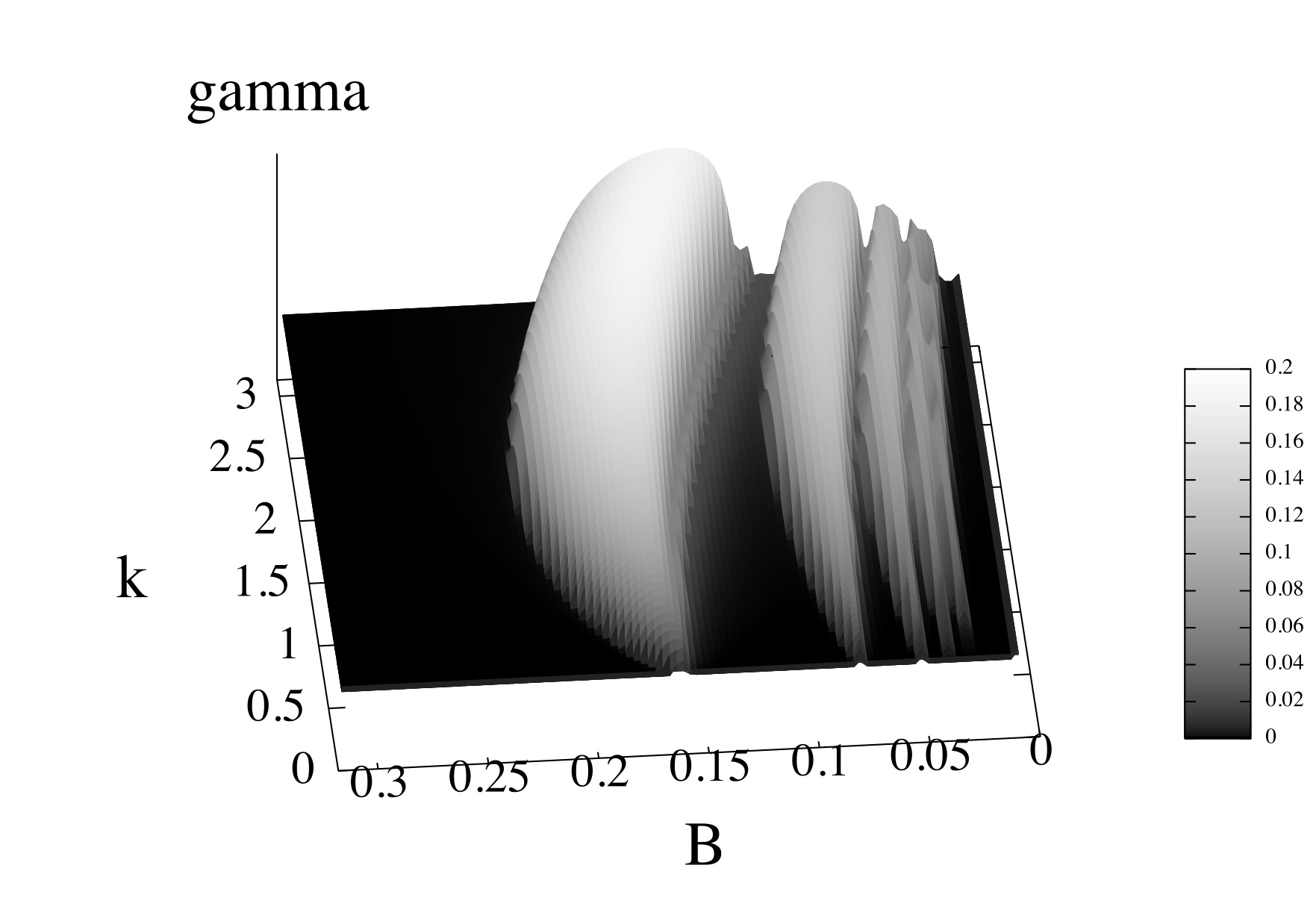}
\end{tabular}
\caption{The FGI  $\gamma$  is plotted for CIC in (a), for LEWIS in (b), and for CELESTE in (c) as a function of $k$ and $B$ for a cold, drifting  plasma.   From left to right, each ridge corresponds   to aliases 1 through 5 of the principal wave number. In (a), the growth rate is zero for all $k$ when $B > 0.25$, and for all $B$ when $k=\pi/\Delta x$. In (b), $\gamma$ is zero for all $k$ when $B>0.4$. In (c), $\gamma$ is zero for all $k$ when  $B>0.2$}  
\label{GammaBCIC}
\end{figure}

The variation of the theoretical growth rate, $\gamma$, with $k \Delta x/2$ and beam 'Debye length', $B=U_0/\omega_p \Delta x$, \cite{birdsall1980},  is shown for CIC, LEWIS and CELESTE in Figs. \ref{GammaBCIC} (a),(b), and (c).  Black shading corresponds to $\gamma=0$, and the white
to the highest values of $\gamma$.   In general, high beam speeds or small values of $\Delta x$ yield stability.  CIC, LEWIS, and CELESTE  are stable for $B > 0.25,  0.4,$ and $ 0.2$ respectively.  The separate contributions of the aliases included in the sums in Eqs. \ref{CICFGI} - \ref{PICeFGI} determine the number of ridges.  From left to right, each higher value of $q$ corresponds to a  ridge with lower height and a smaller cutoff.

CIC is stable to the FGI for all values of $B$ when $k=\pm \pi/\Delta x$ as a consequence of the averaging that defines $E_c$.  The averaging replaces $k_q \Delta x/2$ by $k_q \Delta x$ in Eq. \ref{CICFGI}.  Because $S_k(k_q \Delta x)=0$ for $k=\pi/\Delta x$ and all $q$,  Eq. \ref{CICFGI}  has solutions with real values of $\omega$ only. 
Both LEWIS and CELESTE, Figs. \ref{GammaBCIC}(b) and (c), are unstable to the FGI for $B$ small enough and $k= \pm \pi/\Delta x$.  

The maximum growth rate is smallest for CELESTE, $\gamma_{max}=0.2 \omega_{pe}$,  and largest for LEWIS, $\gamma_{max}=0.4 \omega_{pe}$.  The only difference between LEWIS and CELESTE  is the contribution of the mass matrix.  The modest  smoothing properties of the quadratic b-spline in the charge density calculation in LEWIS and CELESTE are overwhelmed in LEWIS by the anti-smoothing contributed by the mass matrix.

\subsection{Estimated error in CELESTE momentum conservation}

\label{momentum}


It is  shown in Section \ref{momentumconservation} that CIC conserves momentum but CELESTE and LEWIS do not because Gauss' law is not satisfied by $Q_v^{(1)}$ in the total momentum equation.  For CELESTE, a charge density at vertex $v$ computed by averaging $Q_c^{(2))}$, $\tilde{Q}_v=1/2 (Q^{(2)}_c+Q^{(2)}_{c-1})$  satisfies an averaged Gauss'  law, Eq. \ref{GaussCELESTE}, 
$$\tilde{Q}_v = (E_{v+1}-E_{v-1})/2 \Delta x.$$
and the RHS of Eq. \ref{LewisP} is,
$$1/2  \sum_v   (E_{v+1}-E_{v-1}) E_v = \sum_v \tilde{Q_v}  E_v \Delta x,$$
which is zero for periodic boundary conditions and a function of the boundary values otherwise.
Any error in momentum conservation is proportional to the difference between $Q_v^{(1)}$ and $\tilde{Q_v}$.  Since both are computed from the same particle data but use b-splines differently, the analysis to obtain an error estimate is simple.  First, the average $\mathcal{S}^{(l)}$ used in in $\tilde{Q}_v$ is given by, 
$$\frac{1}{2} \left( \mathcal{S}^{(l)}(x_c-x_p) +  \mathcal{S}^{(l)}(x_{c-1}-x_p) \right)=\frac{1}{2}\int \dx \mathcal{S}^{(l-1)} (x-x_v) \Sngp (x-x_v;2 \Delta x), $$
the average of $\mathcal{S}^{(l)}$ at adjacent cell centers is equal to the convolution of $\mathcal{S}^{(l-1)}$ with an $\Sngp$ with support $2 \Delta x$.  Thus, where $q_k \equiv \sum_p q_p \exp(ikx_p) $,
the Fourier transform of the error is,
\begin{equation}
(\epsilon_Q)_k = q_k \mathcal{S}_k^l \left( \frac{sin (k \Delta x)}{k \Delta x}-1 \right ). 
\end{equation} 
For small $|k \Delta x|$, the error scales as $\Delta x^2$, but for modes with $k \Delta x \to \pm \pi$, the error in the charge is as large as $(Q_v)_k =q_k  \mathcal{S}_k^l$.  Any application with deviations from charge neutrality at small scales sufficient to cause particle trapping may have significant errors in momentum conservation.


\subsection{Energy conservation}
\label{NRGerrors}
Here we consider sources of error in energy conservation in the CELESTE and CIC algorithms following the analysis for the implicit moment method in \cite{brackbill1982,vu1992}.  (An analysis of LEWIS gives similar results to CELESTE.)
In Section \ref{ModelNRG}, it is shown that energy conservation means that  the numerical value of the sum of  kinetic and field energies, $\mathcal{K}+\mathcal{F}$, is constant.  Any change is due to error if, as we assume,  periodic boundary conditions apply. 

The change in the particle kinetic energy in a time step, $W_ {\mathcal{K}}$,  is the product of the electric field and  the particle current.  For CELESTE, 
\begin{equation}
W_{ \mathcal{K}}=\sum_v \frac{1}{2} (J_v^{n+1/2}+\widehat{J_v^{n-1/2})}E^n_v \Delta t,
\end{equation}
where the particle currents are,
\begin{equation}
J_v^{n+1/2}=\sum_p q_p v_p^{n+1/2} \Sl (x_v-x_p^n)
\label{current}
\end{equation}
and 
\begin{equation}
\widehat{J_v^{n-1/2}}=\sum_p q_p v_p^{n-1/2} \Sl (x_v-x_p^n).
\end{equation}
The second particle  current , when expanded about position, $x_p^{n-1}$,  becomes 
\begin{equation}
\widehat{J_v^{n-1/2}} = J_v^{n-1/2} - \frac{(\Pi_c^{n+1/2}-\Pi_{c-1}^{n+1/2}) \Delta t}{\Delta x} + ... .
\end{equation}
where the pressure,  $\Pi_c$, is
\begin{equation}
\Pi_c =\sum_p q_p v_p^{n+1/2} v_p^{n+1/2} \Sngp(x_c-x_p^n).
\label{pressure}
\end{equation}
The expansion ends with $\Sngp$.

The changes in particle potential and field energies are equal,
\begin{equation}
\sum_c \left( Q_c^{n+1} \phi_c^{n+1}-Q_c^n \phi_c^n \right) \Delta x = \frac{1}{2} \sum_v (E_v^{n+1})^2-(E_v^n)^2, 
\end{equation}
from which it follows that
\begin{equation}
 \sum_c \left ( Q_c^{n+1}-Q_c^n \right) \frac{1}{2}\left( \phi_c^{n+1}+\phi_c^n \right)=\sum_c Q_c^{n+1} \phi_c^{n+1}- Q_c^{n} \phi_c^{n} .
\end{equation}
The particles move and the charge density,
Eq. \ref{Step1Lewis},  changes from one time step to the next,
\begin{equation}
Q_c^{(l)n+1}-Q_c^{(l)n}=\sum_p q_p \left( \Sn(x_p^{n+1}-x_c)- \Sn(x_p^n-x_c) \right).
\end{equation}
For CELESTE with $l=2$,  expand $\Sq$ about $x_p^n$ to derive,
\begin{equation}
Q_c^{n+1}-Q_c^n=-\frac{(\widetilde{J_{v+1}^{n+1/2}}-\widetilde{J_v^{n+1/2}})\Delta t}{\Delta x} +O((v_p \Delta t/\Delta x)^3,
\end{equation}
where $\widetilde{J}$ includes contributions from the pressure,
\begin{equation}
\widetilde{J_v^{n+1/2}} \equiv J_v^{n+1/2}- \frac{\left( \Pi_c^{n+1/2}-\Pi_{c-1}^{n+1/2}\right ) \Delta t}{2 \Delta x}.
\label{Qcontinuity}
\end{equation}
The expansion is valid only while a particles remain within the support of $\Sq$ \cite{hockney1971}. (This limit is overcome in \cite{chen2011}, where particle orbits are computed in segments bounded by cell boundaries.  Energy is conserved even for $v_p \Delta t > \Delta x$.)



Summation by parts yields the change in field energy, which, except for the first and last time steps, can be written.
\begin{equation}
W_{ \mathcal{F}}=- \sum_v    \frac{1}{2} E_v^{n} \left(\widetilde{J_v^{n+1/2}}+ \widetilde{J_v^{n-1/2}} \right) {\Delta t} . 
\label{CELESTEW_F}
\end{equation}
Combining the kinetic and field energy changes gives us the energy error, $\Delta \mathcal{E}=W_{ \mathcal{K}}+W_{\mathcal{F}}$, per time step for CELESTE,
\begin{equation}
\Delta \mathcal{E}=\sum_v E_v^n \left(\frac{1}{2} (J_v^{n+1/2}+\widehat{J_v^{n-1/2}}) - \frac{1}{2} (\widetilde{J_v^{n+1/2}}+\widetilde{J_v^{n-1/2}}) \right ) \Delta t+O(v_p \Delta t)^3.
\end{equation}
The surprising result is that the pressure terms cancel and the   error in  CELESTE energy is $O((v_p \Delta t)^3$, and the error scaling is given by,
\begin{equation}
\Delta \mathcal{E} \approx \beta' (v_p \Delta t / \Delta x)^3.
\label{no-avg}
\end{equation}

For CIC, $W_{\mathcal{K}}$ with $E_c=1/2(E_{v+1}+E_v)$ interpolated to the particles is,
\begin{equation}
W_{\mathcal{K}}=\sum_c \frac{1}{2}(J_c^{n+1/2}+ \widehat{J_c^{n-1/2}}) \frac{1}{2} (E^n_{v+1}+E^n_{v}) \Delta t,
\end{equation}
where
$$J_c^{n+1/2}=\sum_p q_p v_p^{n+1/2} \Sl(x_c-x_p^n),$$
and
$$\widehat{J_c^{n-1/2}}=\sum_p q_p v_p^{n-1/2} \Sl(x_c-x_p^n).$$
When $W_{\mathcal{K}}$ is re-summed over $v$, it can be compared with  CELESTE, Eq. \ref{CELESTEW_F},
\begin{equation}
W_{\mathcal{K}}=\sum_v \frac{1}{2} \left( \left \langle J_v^{n+1/2}\right \rangle+ \left \langle \widehat{J_v^{n-1/2}} \right \rangle \right ) E^n_{v} \Delta t,
\end{equation}
where, 
\begin{equation}
 \left \langle J_v\right \rangle= \frac{1}{2} \left( J_c +J_{c-1} \right).
\end{equation}

The charge continuity equation for CIC,
\begin{equation}
Q_c^{n+1}-Q_c^n=-\frac{({J0_{v+1}^{n+1/2}}-{J0_v^{n+1/2}})\Delta t}{\Delta x} +O((v_p \Delta t/\Delta x)^2,
\end{equation}
where 
$$J0_v^{n+1/2}=\sum_p q_p v_p^{n+1/2} \Sngp(x_v-x_p^n),$$
contains no pressure contribution because the expansion ends with $\Sngp$.

$ \left \langle J_v \right \rangle$ and  $J0_v $ are not equal in the CIC error estimate, 
\begin{equation}
\Delta \mathcal{E}=\sum_v E_v^n \left[ \frac{1}{2} \left ( \left \langle J_v^{n+1/2}\right \rangle+ \left \langle J_v^{n-1/2} \right \rangle \right ) - \frac{1}{2} (J0_v^{n+1/2}+J0_v^{n-1/2} \right ] \Delta t,
\end{equation}
because of averaging and the absence of a pressure contribution in the charge continuity equation, so the error is different for CIC and CELESTE.
To estimate the error due to averaging,  we use the Fourier transform of the term in brackets,
$$ J_k=j_k \mathcal{S}_k \left ( \frac{sin(k \Delta x)}{k \Delta x}-1 \right) \approx - j_k \mathcal{S}_k \frac{(k \Delta x)^2}{6},$$
where $j_k=\sum_p  exp(-i k x_p) q_p v_p .$
The error scaling for CIC is approximated by,
\begin{equation}
\Delta \mathcal{E} \approx \alpha \Delta x^2 + \beta (v_p \Delta t/\Delta x)^2,
\label{averaging}
\end{equation}
where the undetermined constants $\alpha$ and $\beta$ multiply the error contributed by averaging and  by the charge continuity equation without pressure respectively.

\section{Numerical Test Problems}
The numerical parameters one can vary are the number of simulation particles, $N_p$,   the time step, $\Delta t$, and the grid resolution, $\Delta x$.  The numerical and physical parameters can be linked by the ratio of the Debye length to the cell size, 
\begin{equation}
D=\lambda_D/\Delta x,
\label{D}
\end{equation}
 the  time step in units of inverse plasma frequency, 
 \begin{equation}
 \Delta T=\omega_{pe} \Delta t,
 \label{dT}
 \end{equation}
 and the number of cells a particle travels at the thermal speed in a time step,
  \begin{equation}
  C=D \Delta T=v_{te} \Delta t/\Delta x.
  \label{C}
  \end{equation}  

When the dimensionless parameters, $D, \Delta  T,$ and  $C$  have a nominal value of 1, then $\Delta x$ and $\Delta t$ resolve  $\lambda_D$, $\omega_{pe}$, and the particle motion.  This is essentially  the guidance given to users for the choice of numerical parameters in  \cite{birdsall1985}.  In addition, it is remarked that the number of simulation particles  per Debye length, $$N_D=N_p \lambda_D/L,$$ should be significantly larger than 1 to control numerical fluctuations in charge \cite{birdsall1985}.

The total energy  and momentum, $\mathcal{E}$ and $\mathcal{P}$, are constants of the motion for all the problems we consider.  Any changes in these over time are due to numerical error.  We define an energy error, $\epsilon$, as the per cent change in the energy from $T=0$ to $T=T_{final}$, $$\epsilon\equiv100\times \frac{ \mathcal{E}(T_{final})-\mathcal{E}(0)}{\mathcal{E}(0)}.$$

\subsection{Buneman instability}

The Buneman instability describes the interaction of drifting electrons with initially stationary ions that  results in exponential growth of an electrostatic potential.  The potential  traps thermalize the electrons
and transfers their momentum to the ions.  
CELESTE and CIC simulations of the Buneman instability agree with each other and with  theory. The numerical parameters are such that for both CELESTE and CIC  errors in the conservation of momentum and energy are small.

The case considered in \cite{hirose1982} is reproduced here.  With a periodic domain, $L=c/\omega_{pe}$, (just one electron collisionless skin depth) electrons with drift speed $U_e =0.159 c$ and thermal speed, $v_{te}=0.02 c$ stream through stationary ions with $v_{ti}=0.001 c$.  The mass ratio, $m_i/m_e=100$.  

Excellent energy and momentum conservation is realized by both CIC and CELESTE simulations on
a computational grid with 128 cells and 500 simulation particles of each species in each cell to model the plasma.  With these parameters,  $N_D=64000$ and $D =20$.  The simulations are advanced with constant time step $ \Delta T =0.25$ using an explicit leapfrog algorithm.  The CIC results are reported in \cite{hirose1982}.  The CELESTE results are given here.  

\begin{figure}
\begin{tabular}{cc}
(a) &  (b)  \\
\includegraphics[width=60mm]{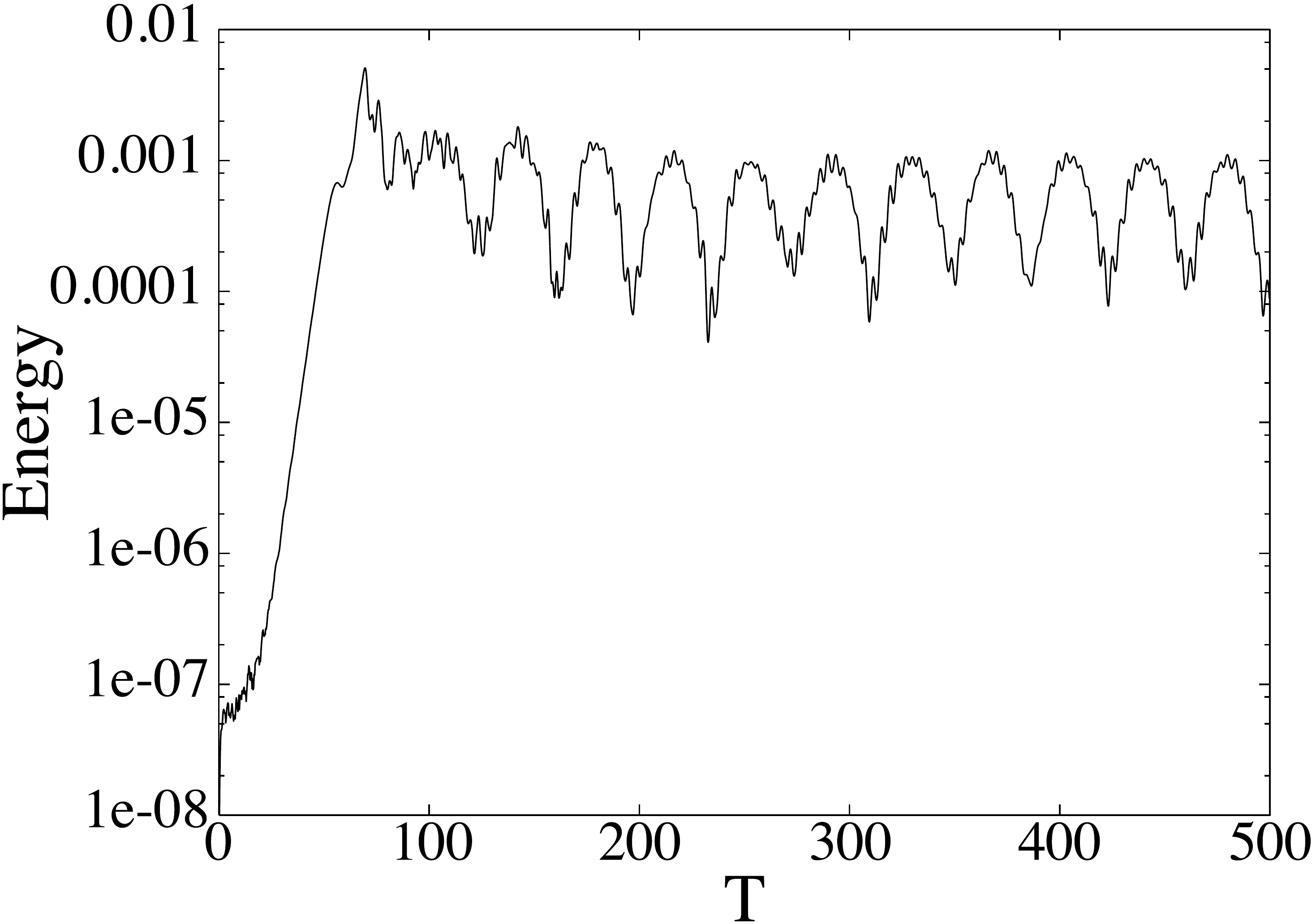} &
\includegraphics[width=60mm]{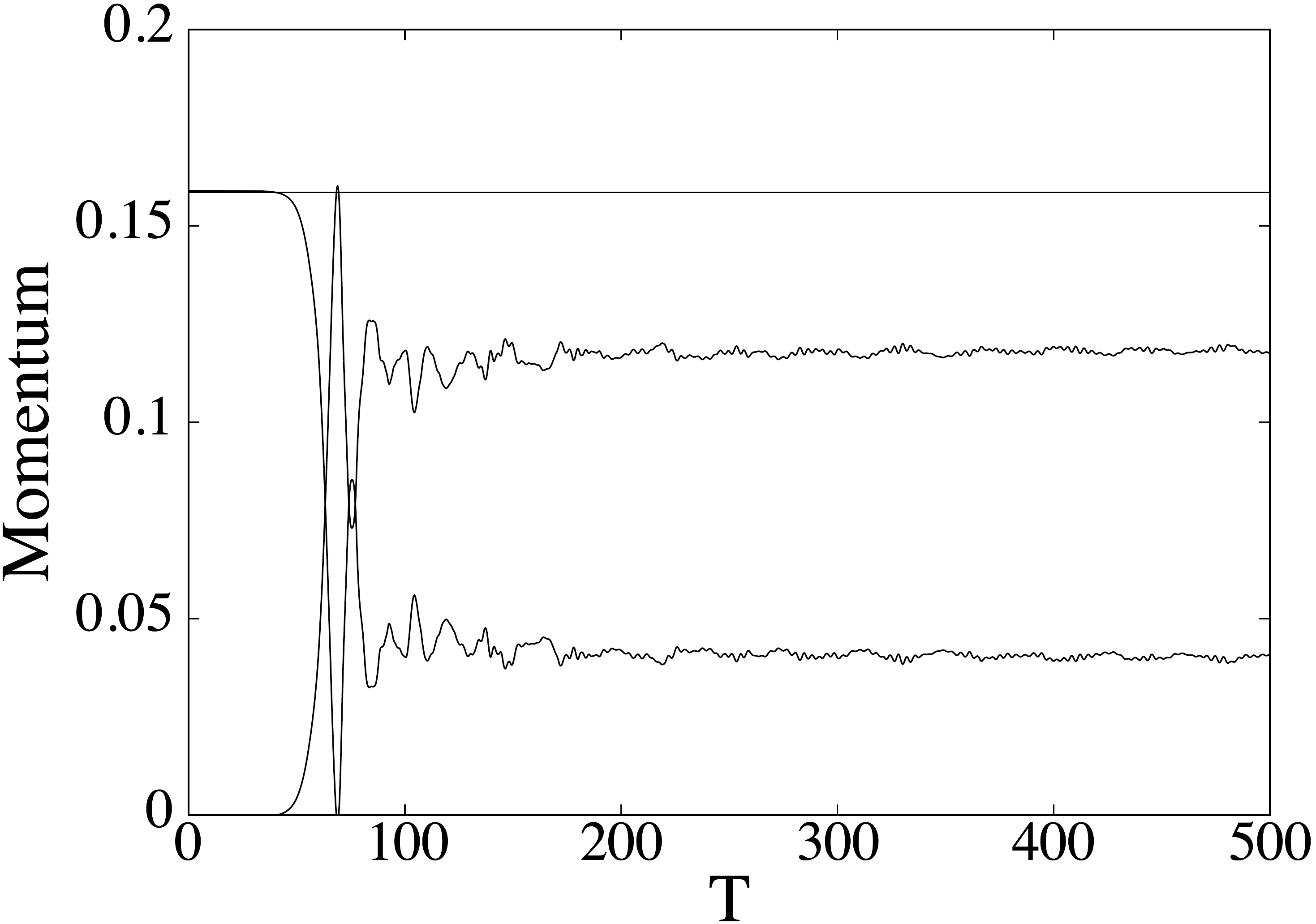}  \\
(c)  &  (d)  \\
\includegraphics[width=60mm]{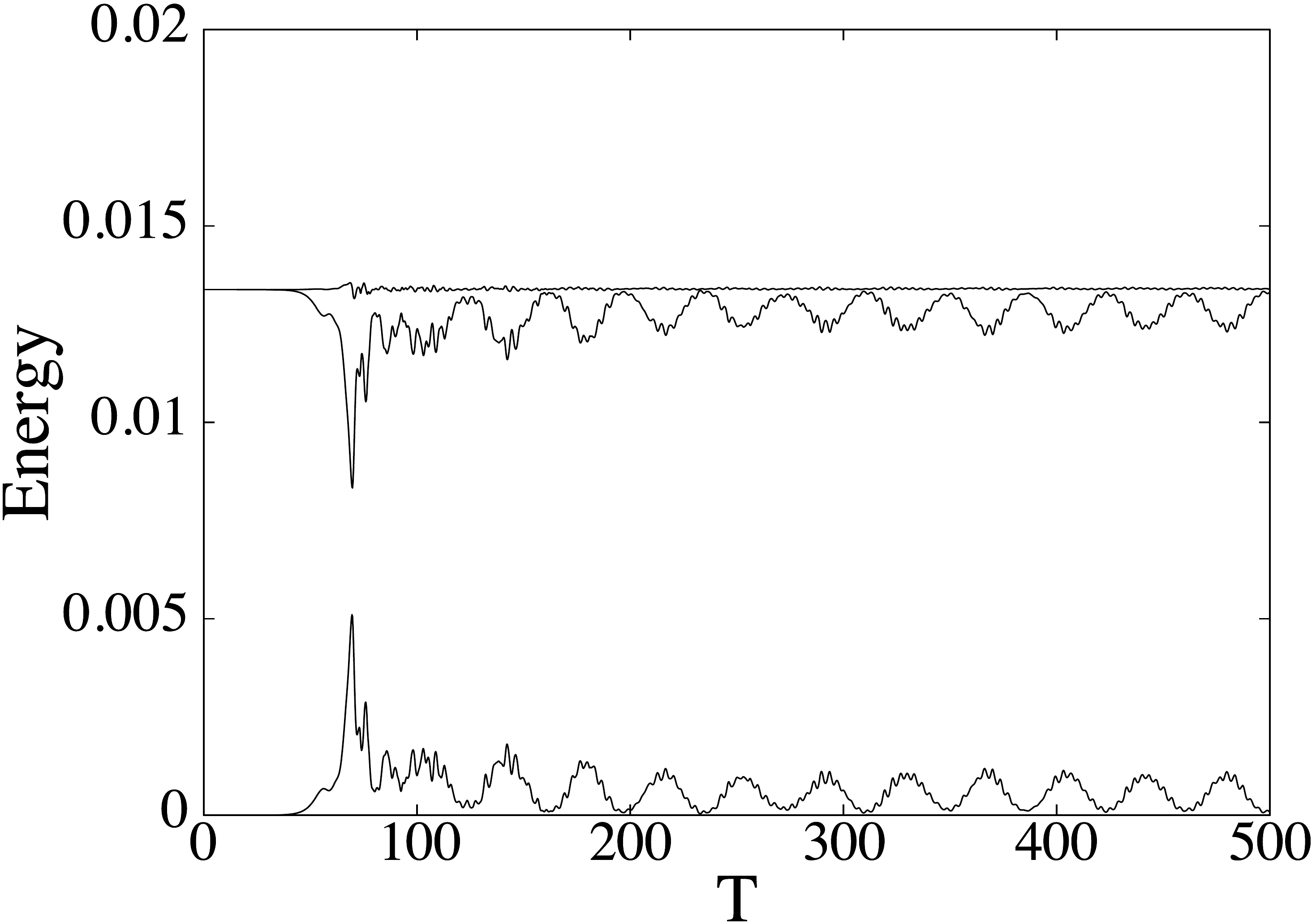} &
\includegraphics[width=60mm]{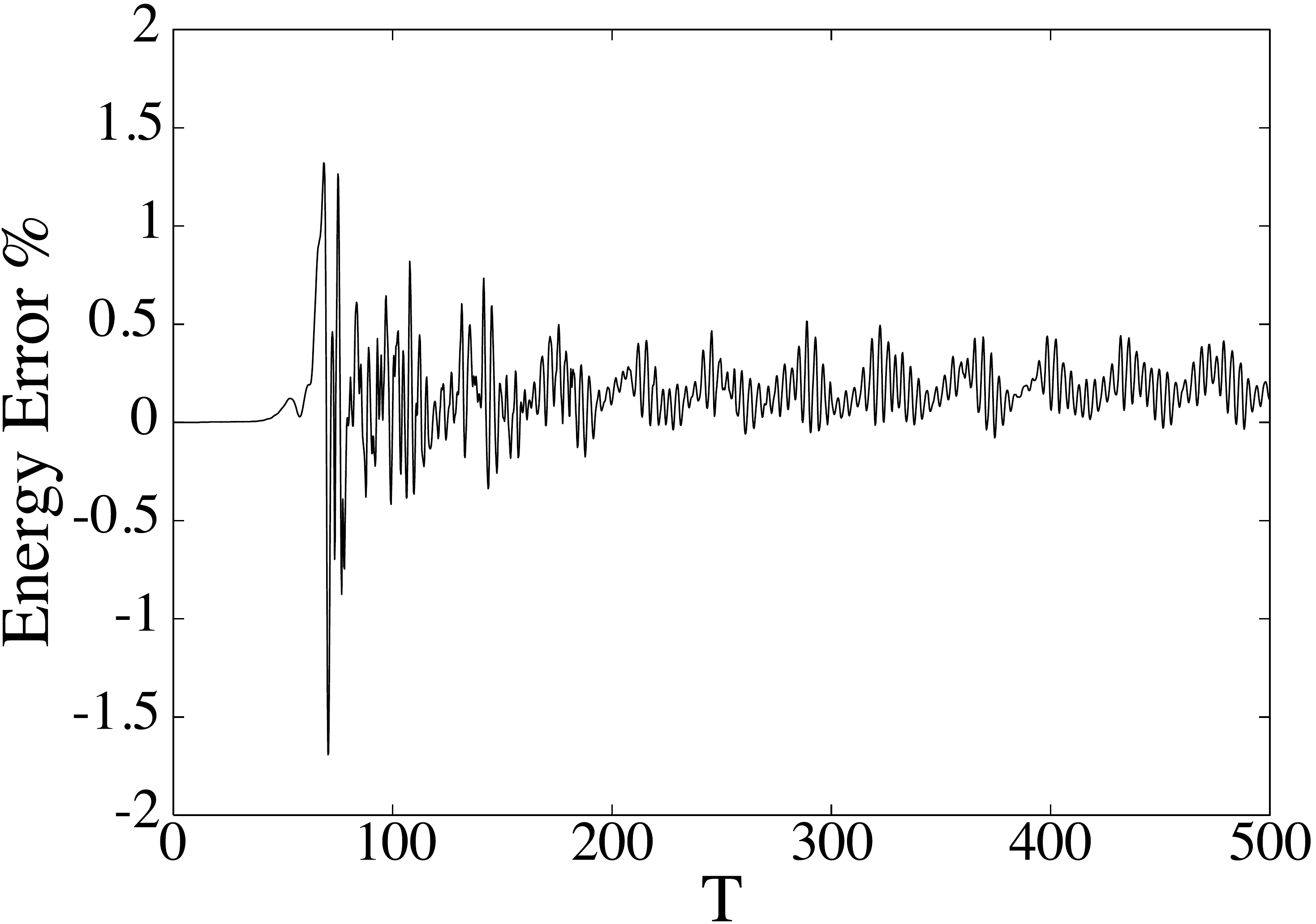}
\end{tabular}

\caption{Results are shown from an energy-conserving simulation of the Buneman instability with $m_i/m_e=0.01$, $D=2.56$, and $B=20.35$:   (a) The electric field energy's unstable growth saturates due to particle trapping. (b) Momentum is exchanged between electrons and ions while total momentum is constant. (c) The maximum in  field energy and the minimum in kinetic energy at $T=70$ add to give constant total energy. (d)     }
\label{ENRG}
\end{figure}

The exponential growth of the field energy, Fig. \ref{ENRG}(a),  is terminated by particle trapping and followed by persistent
ion acoustic oscillations, 
Total momentum is conserved, but particle trapping causes a transfer of momentum from the drifting electrons to the initially stationary ions, Fig. \ref{ENRG}(b).  The trapping time is short compared with the time required for the field energy to grow.
The gain in field energy is balanced by a loss in kinetic energy, Fig \ref{ENRG}(c).  Errors in the total energy are less than $1 \%$ for CELESTE, Fig.  \ref{ENRG} (d). 

\subsection{Cold drifting plasma}

Different results from those in \cite{hirose1982} are described by Okuda \cite{okuda1972} and Langdon \cite{langdon1973} for a cold beam, 
$v_{te}=0.0$,  $U_0=0.16c$, and stationary ions.  Simulations with several algorithms, including CIC and LEWIS with $l=1$, are performed on a computational grid with 64 cells, domain $x \in [0, 64 c/\omega_{pe}]$, and $16$ particles per cell to represent the electrons.   The ions are stationary,  $m_e/m_i=0$, and there is no Buneman instability because the instability growth rate scales as $(m_e/m_i)^{1/4}$.  There is, instead, an FGI, Section \ref{aliasing}. The ratio of the Debye length to the grid spacing is
$D=0.16$, compared with $D=20$ in \cite{hirose1982}.

\begin{figure}
\begin{center}
\begin{tabular}{cc}
(a)  &  (b)  \\
\includegraphics[width=60mm]{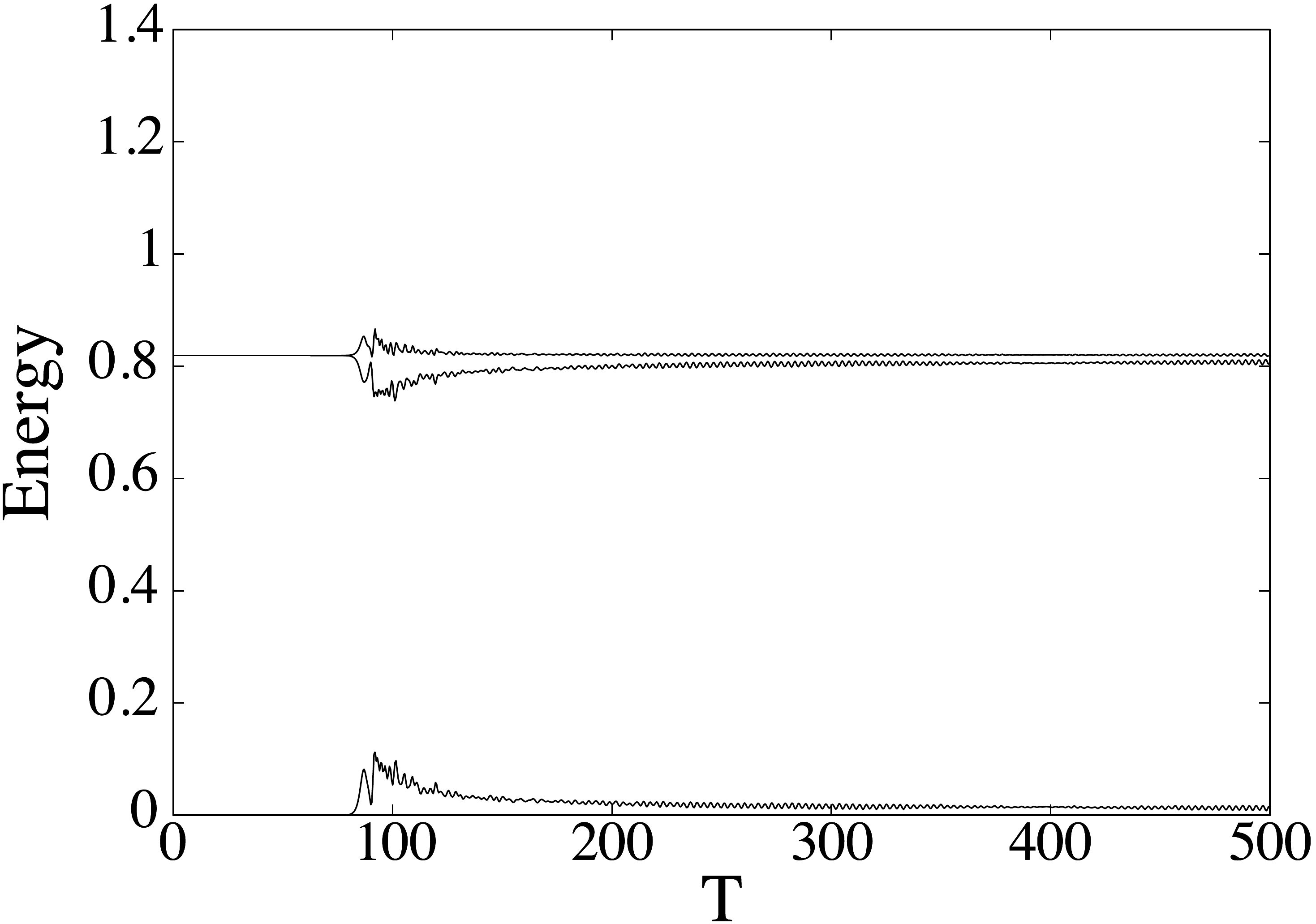}  &
\includegraphics[width=60mm]{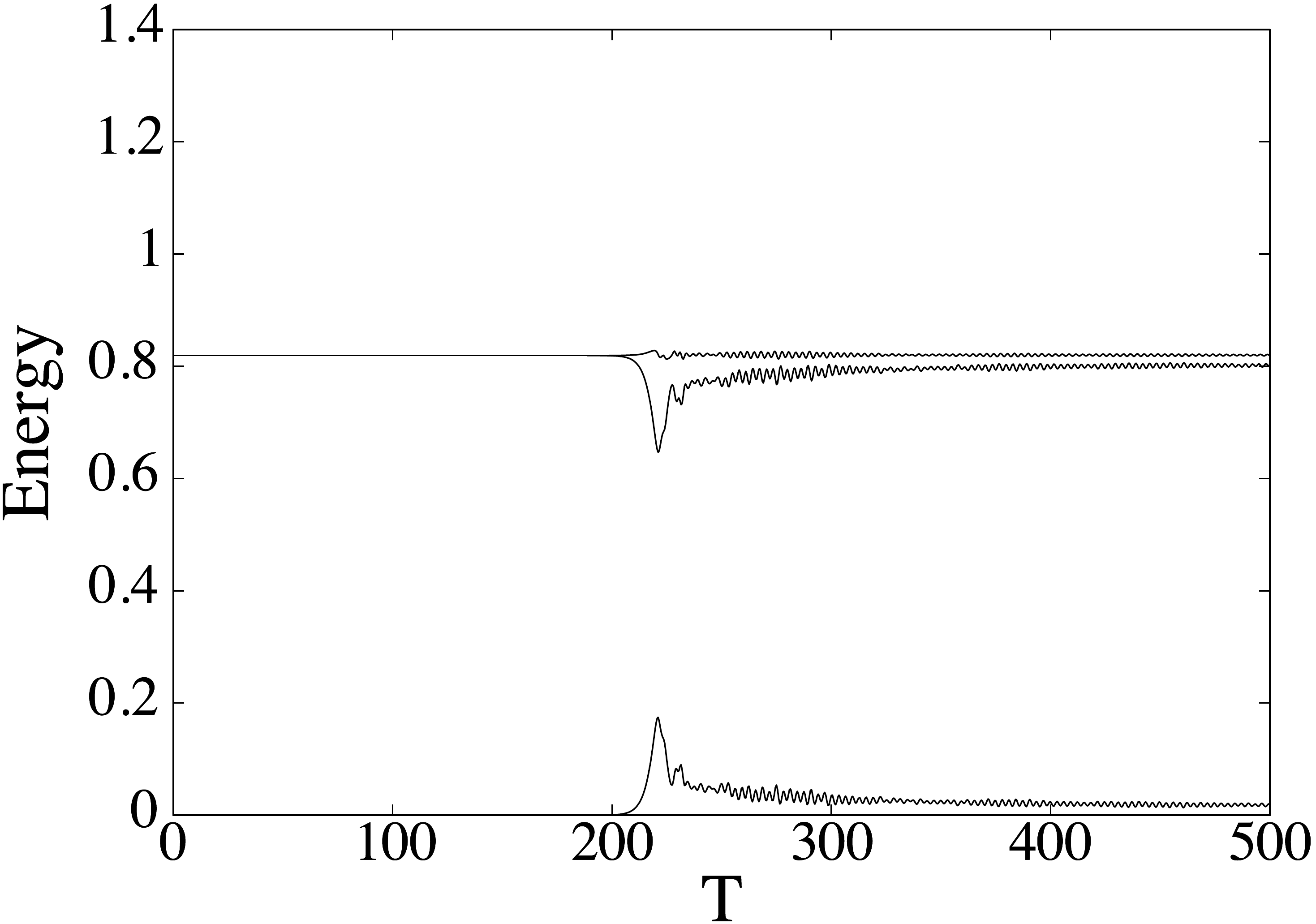} 
\end{tabular}
\end{center}
\caption{ For both LEWIS (a) CELESTE (b), variation in $\mathcal{E}$ (upper curve) is smaller than changes in $\mathcal{F}$ (lower curve) and $mathcal{K}$ (middle curve).    }
 \label{NRGLEWIS}
\end{figure}

First, we compare the electrostatic energy growth computed by CIC, LEWIS (with $l=2$), and CELESTE,  Fig. \ref{NRGLEWIS}.
The instability grows from round-off after a brief latent period.  CELESTE and CIC have similar growth rates, but LEWIS is higher.  The linear growth rate for CIC, LEWIS, and CELESTE is $\gamma/\omega_{pe}=0.105, 0.2$,  and $0.075$ respectively.  The peak CIC value of $E^2$ is more than $400 \%$ and LEWIS is $85 \%$ that for CELESTE.

\begin{figure}

\includegraphics[width=120mm]{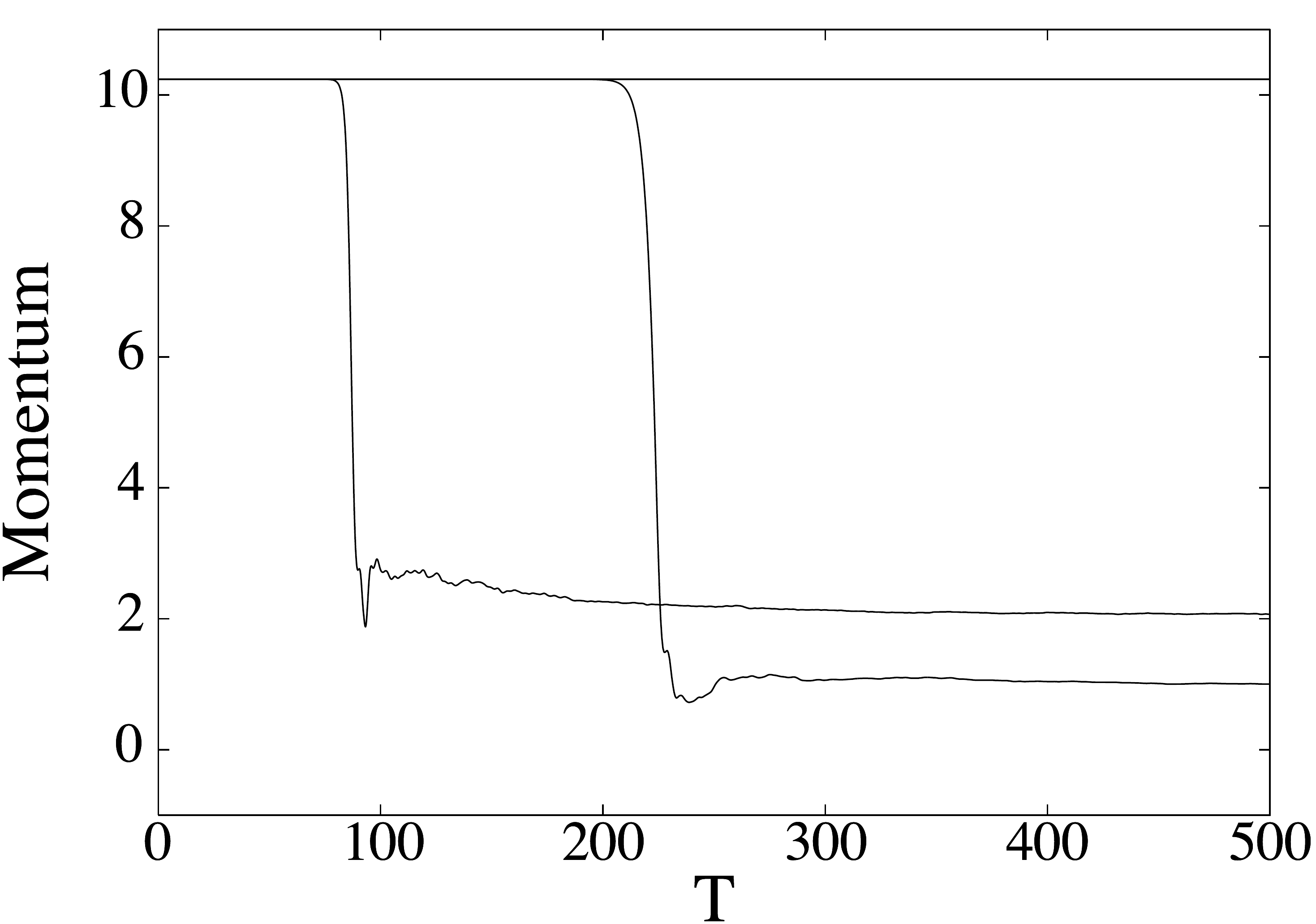} \\

\caption{ The momentum is constant in CIC simulations of a cold beam (upper curve), but CELESTE and LEWIS simulations, electron momentum is lost due trapping in potential wells created by the FGI.  Trapping occurs earlier with LEWIS but the momentum loss is larger  with CELESTE.   }
 \label{Momentum}
\end{figure}

Next, we compare momentum, which is conserved by CIC but not by CELESTE or LEWIS,  Fig. \ref{Momentum}. 
The upper curve at a constant value of $10$ is the result from a CIC simulation.  The two lower curves are the result of  CELESTE and LEWIS simulations, in which nearly all the initial electron momentum is lost  just as the growth of the field energy saturates.  Particle trapping causes saturation of the FGI earlier in LEWIS than in CELESTE, but the  loss of electron momentum with LEWIS is slightly less.  Like the Buneman instability \cite{hirose1982}, the rapid loss is caused by electron trapping in short wavelength  potential wells.  Unlike  \cite{hirose1982}, the instability is numerical and momentum is lost to the computation grid.  

 \begin{figure}
 \includegraphics[width=100mm]{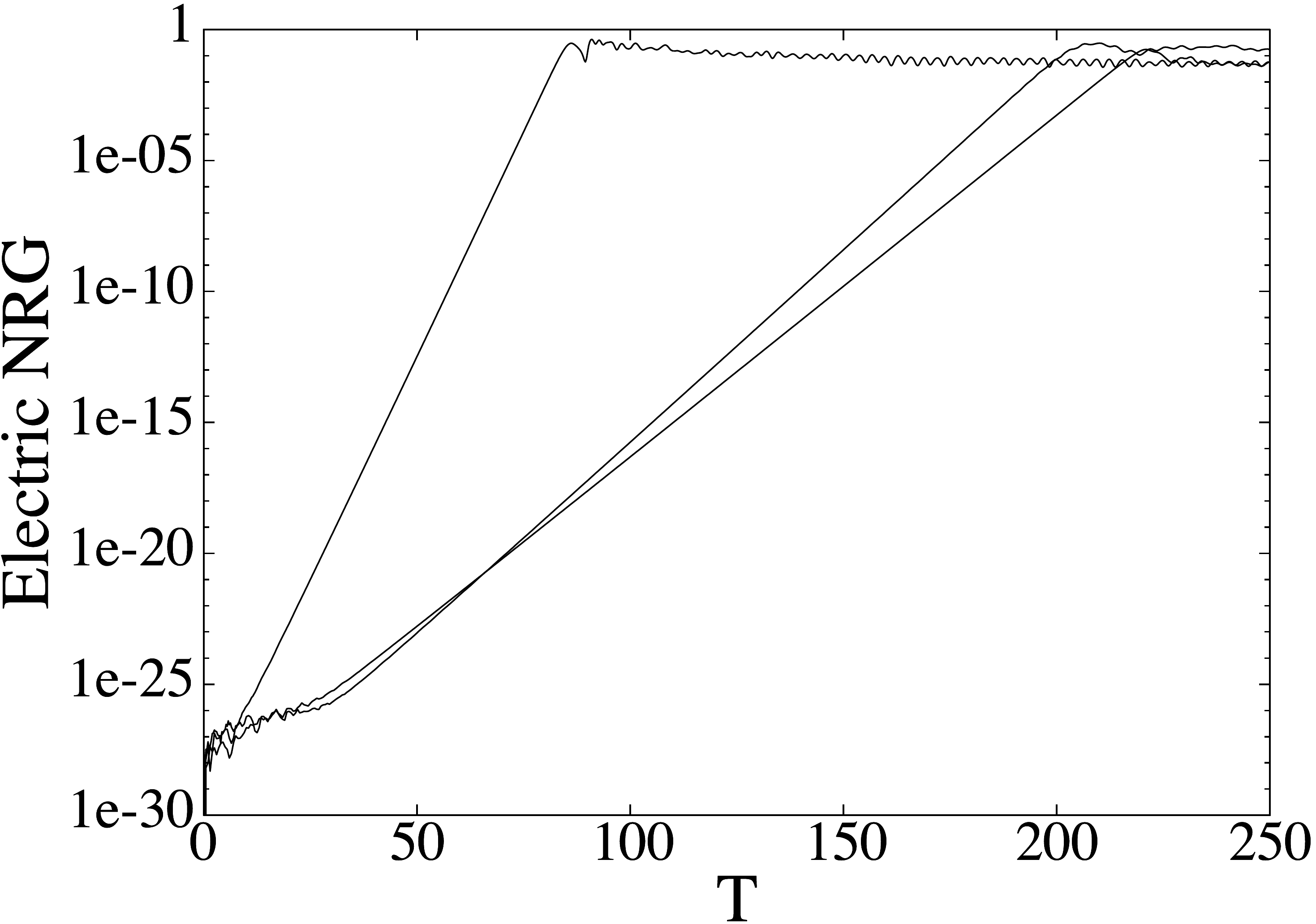}
 \caption From left to right (largest to smallest FGI $\gamma$ for a cold beam),$\mathcal{F}$, is plotted for LEWIS, CIC and CELESTE. 
 \label{NRGELEWIS}
 \end{figure}

\begin{figure}
\includegraphics[width=120mm]{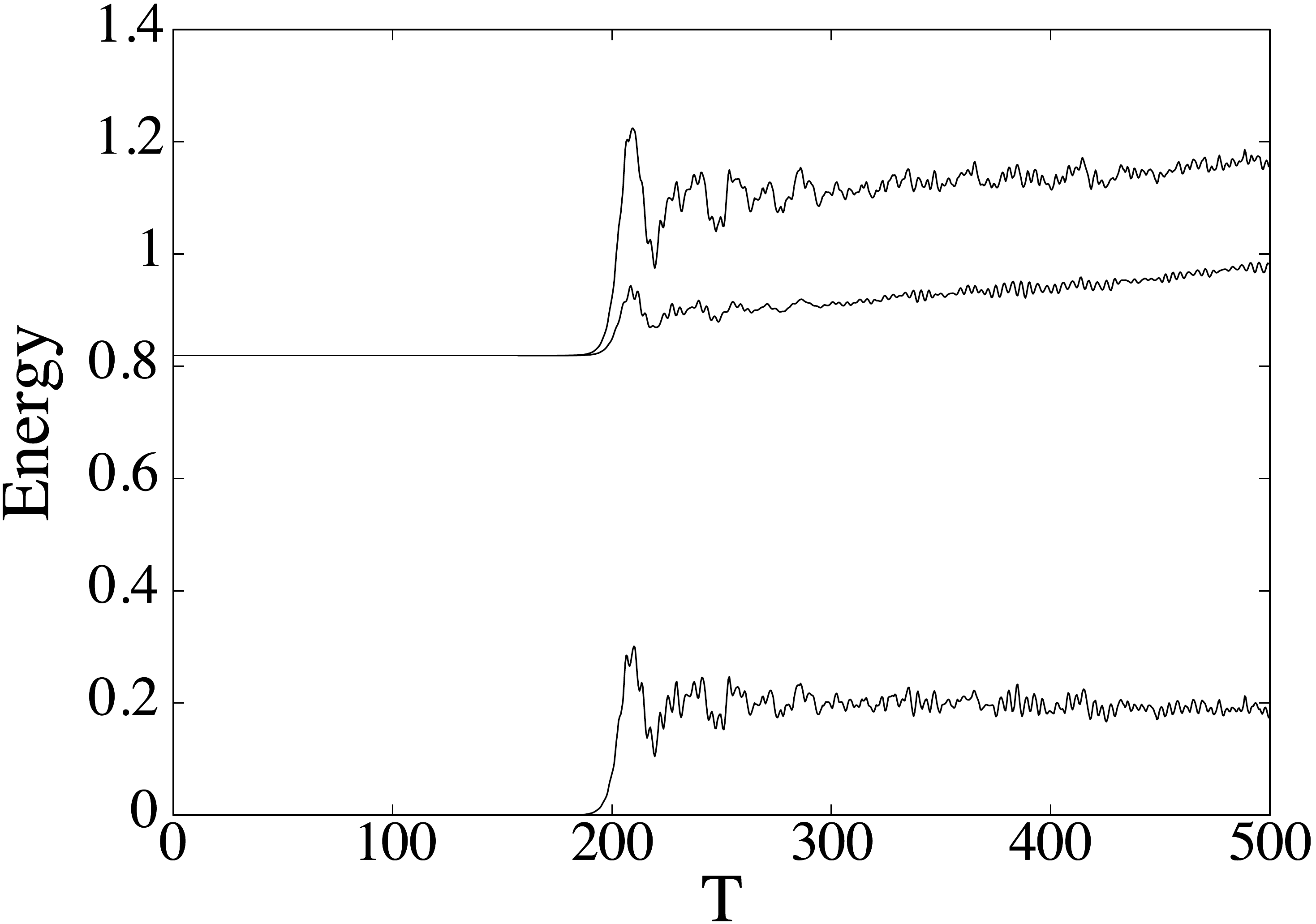} 
\caption{ The CIC $\mathcal{E}$ (upper curve) increases by 40 $\%$ due to the growth in $\mathcal{F}$, bottom curve.   Subsequently,  increases in $\mathcal{K}$ (middle curve) reflect heating.  }
 \label{NRGCIC}
\end{figure}

Energy is conserved by LEWIS and CELESTE, Figs. \ref{NRGLEWIS}(a) and (b) but not by CIC, Fig. \ref{NRGCIC}, for which energy increases by more than $100 \%$ by $T=500.$   The potential energy is constant after saturation, and the drift  energy is small.   Electron heating accounts for almost all the increase in energy with CIC.

\begin{figure}
\begin{tabular}{cc}
  (a)    &   (b)  \\
\includegraphics[width=60mm]{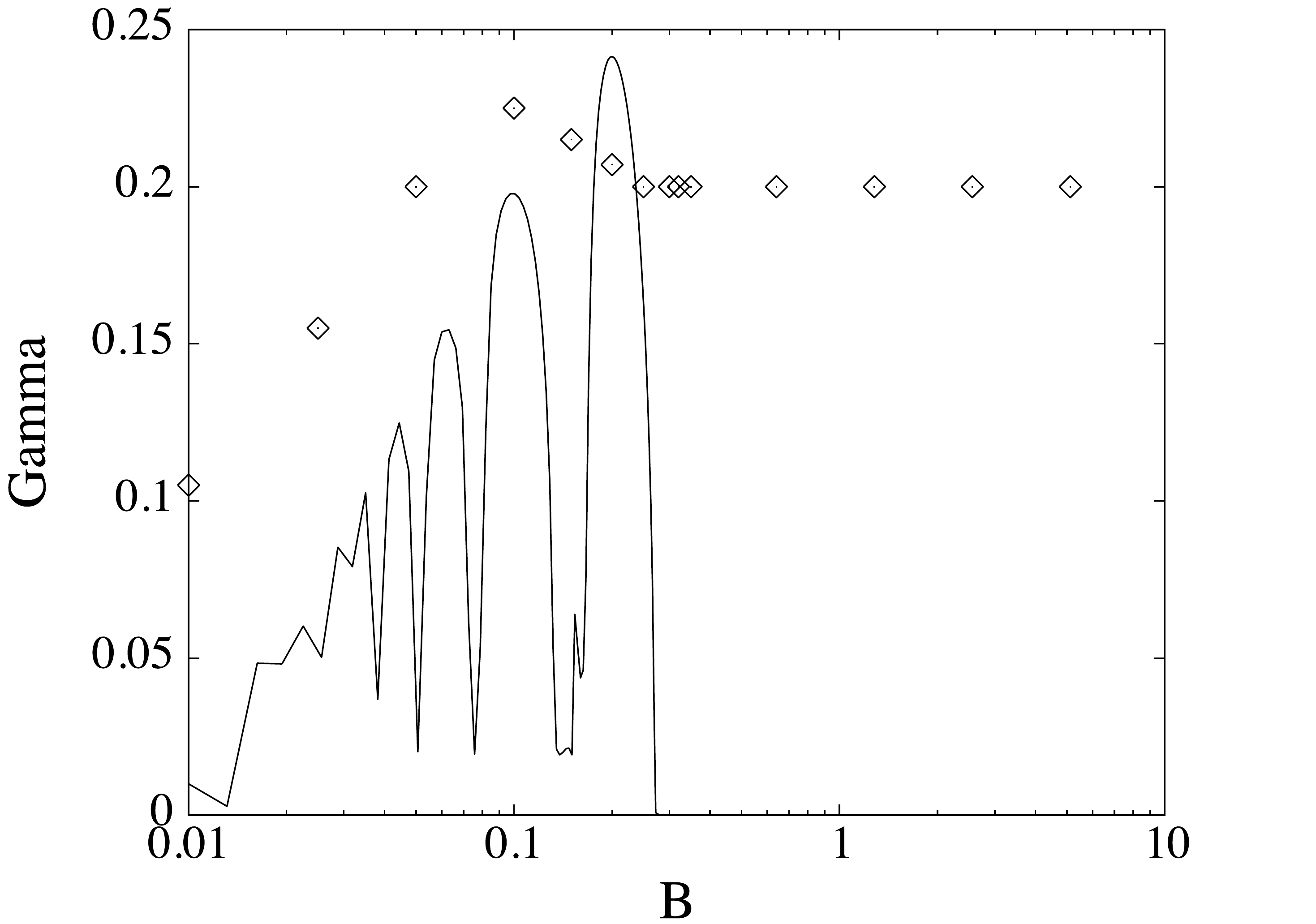}   &
\includegraphics[width=60mm]{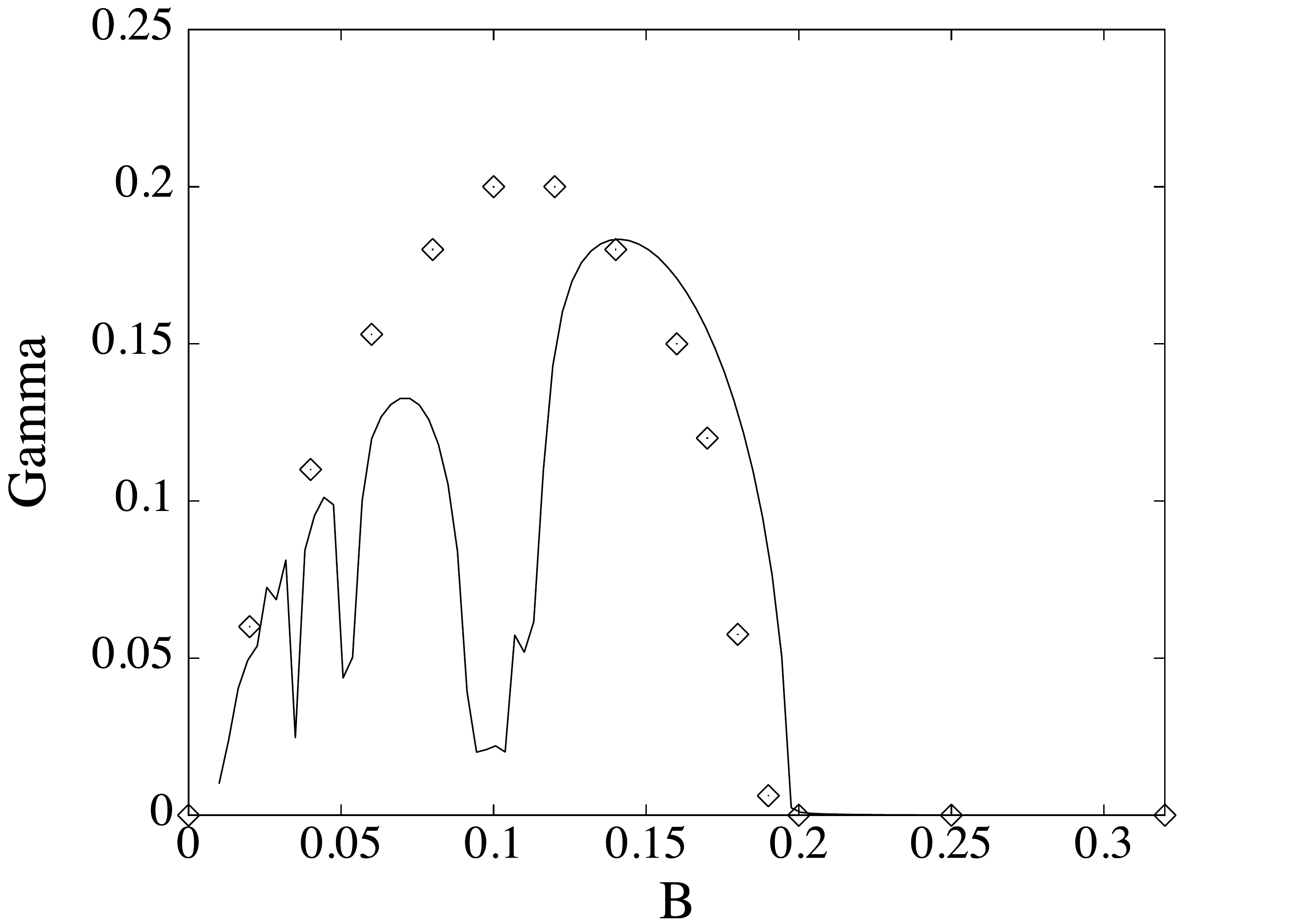} 
\end{tabular}
\caption{The theoretical growth rates for the FGI are compared with numerical results ( $\diamond$'s); in (a) for CIC and in (b) for CELESTE.} 
\label{CICGamma}
\end{figure}

The value of $\gamma$ for CIC and CELESTE from theory and numerical simulations is compared in Fig. \ref{CICGamma}.
A theoretical growth rate is plotted for $k=\pi/2 \Delta x$, which corresponds approximately to the position of the maximum growth rate.  The $\gamma$ from numerical results is the measured growth rate of the electrostatic energy.  

Where the  CIC dispersion theory predicts stability for $B>0.25$, the numerical results exhibit instability, Fig. \ref{CICGamma}(a).  In fact, for all $B >0.25$ the numerical results for CIC are unstable  
with constant growth rate, $\gamma \approx 0.2$, independent of $B$ and not much less than its maximum value. Similar results are described in \cite{birdsall1980}.

By contrast, the CELESTE compare well with dispersion theory in Fig. \ref{CICGamma}(b).  The numerical simulations are stable for $B>0.2$ and the maximum growth rate is $\gamma \approx 0.2$ as theory predicts, However, the separate contributions of the aliases are not visible in the simulation results.

\begin{figure}

\includegraphics[width=120mm]{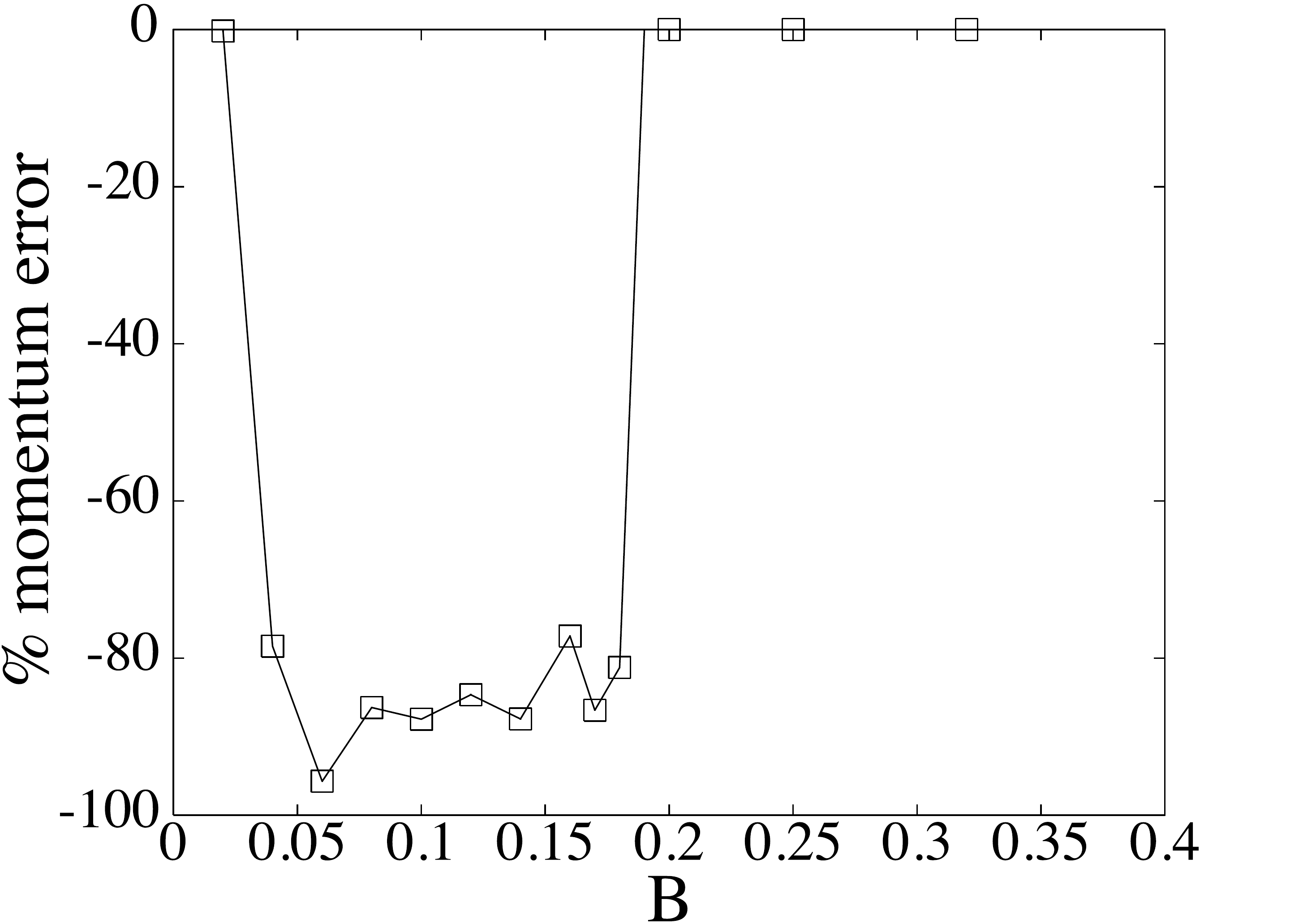}
\caption{ For $B>0.2$, electrons lose no momentum in CELESTE simulations of a cold beam.  }
\label{PvsB}
\end{figure}

Errors in momentum conservation from cold beam simulations with CELESTE are shown in Fig. \ref{PvsB} for $0.02 \leq B \leq 0.32$.   For $0.02 < B<0.2$ a cold beam loses almost all momentum by $T=500$ .  For $B>0.2$, a cold beam loses none of its momentum even when the simulation is continued to $T=10000$.   For values of $B$ for which there is instability, Fig.\ref{CICGamma}( b), there is momentum loss.  Where there is no instability, there is no momentum loss.   In practical terms, if $\Delta x$ is sufficiently small  that $B=\lambda_{Debye}/\Delta x > 0.2$ ,  CELESTE simulations of a cold beam conserve both momentum and energy.

\begin{figure}
\begin{tabular}{cc}
(a)  &  (b)  \\
\includegraphics[width=60mm]{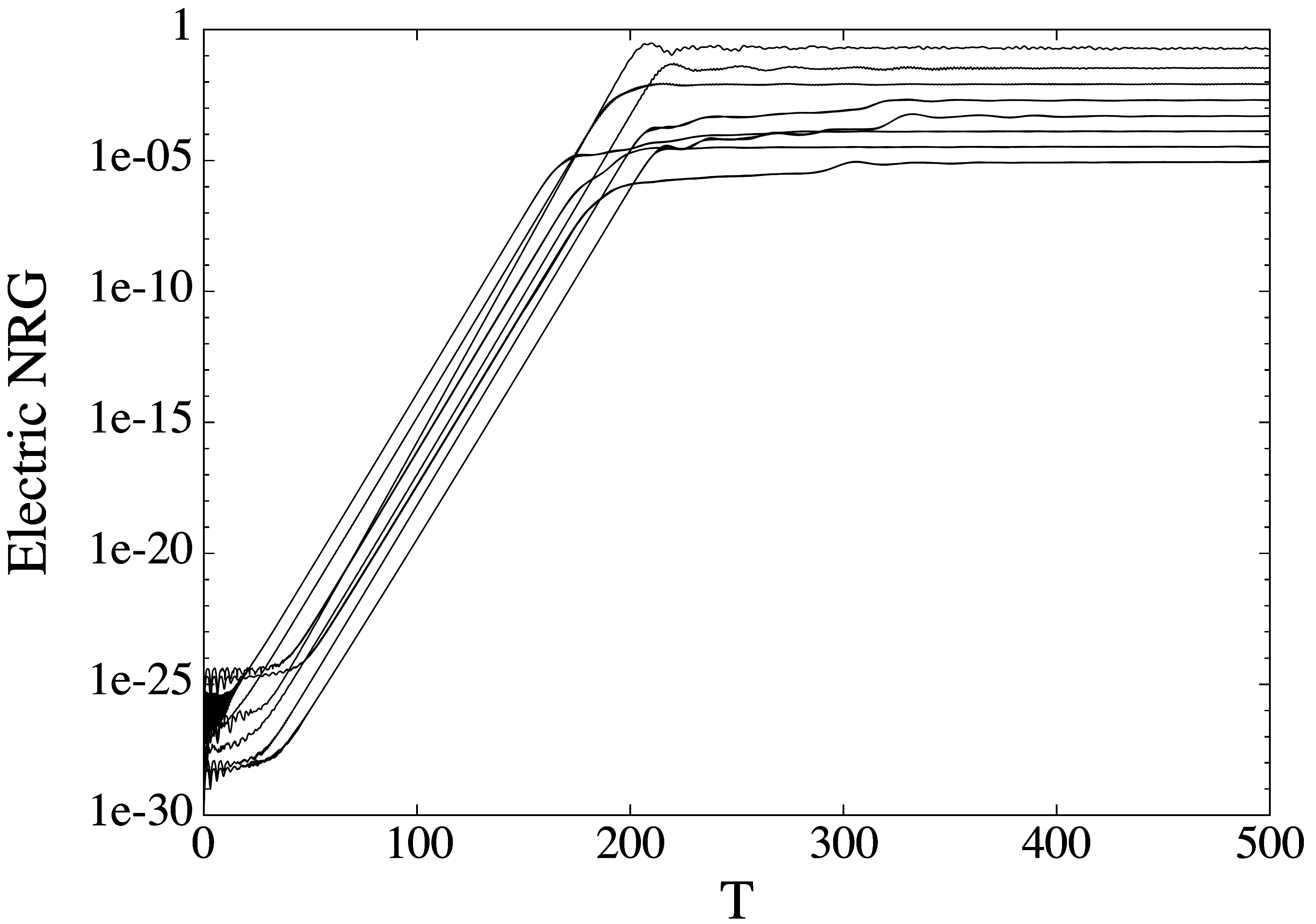}  &
\includegraphics[width=60mm]{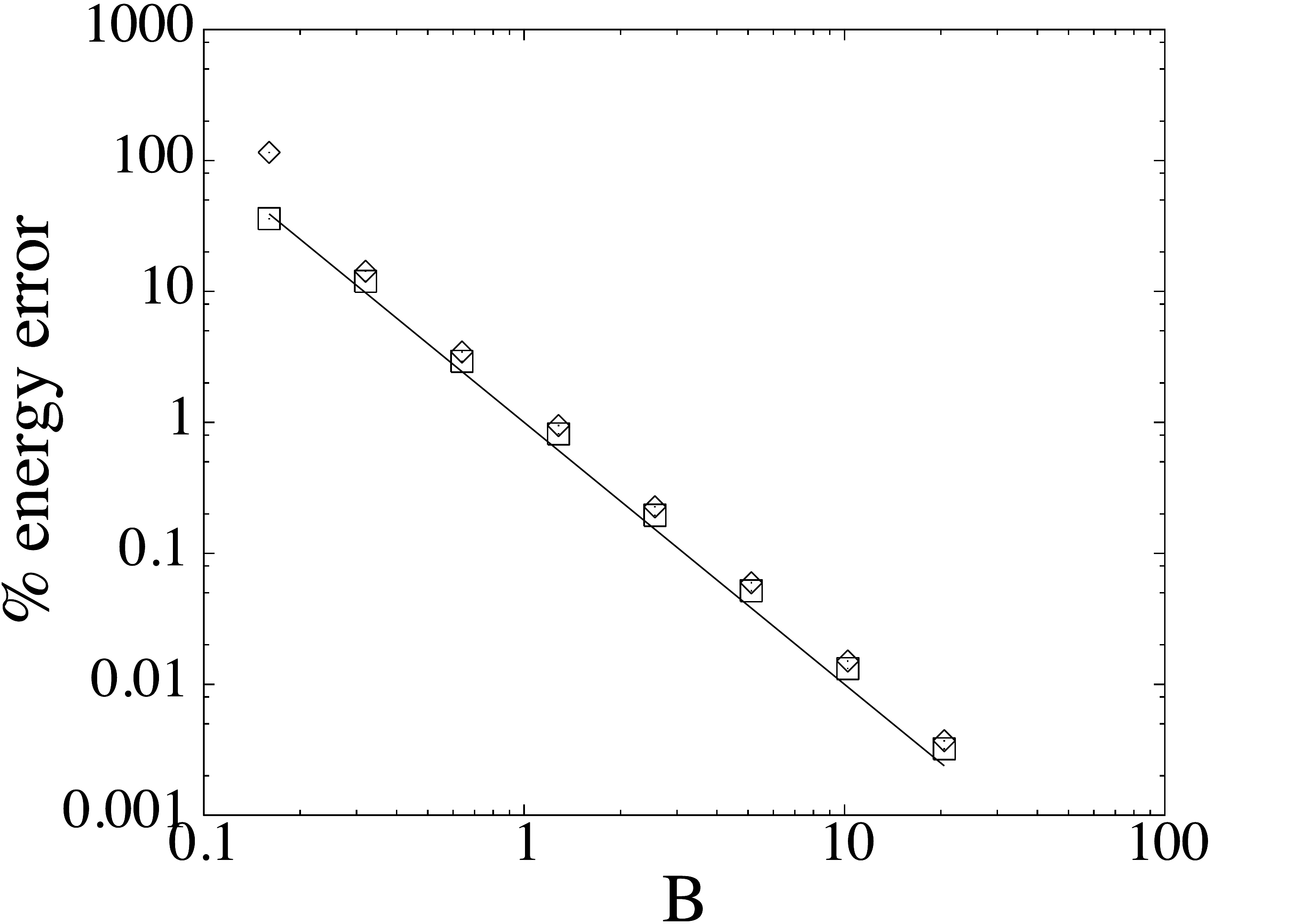} 
\end{tabular}
\caption{  $\mathcal{F}$ from CIC simulations of a cold, drifting plasma with constant $C=0.08$ and  $B$ from Eq. \ref{Bsequence} are compared.  (a)  $\gamma=0.2 \omega_{pe}$ for all values of $B$, but the  saturation energy varies as $1/B^2$ (b) $\mathcal{F}$ at $T=500$ with $C=0.08$, $\Box$'s, is compared with the energy error, $\diamond$'s, and $f(B)=1/B^2$ for reference.   } 
\label{NRGEvsT}
\end{figure}

 The CIC simulation results in Fig. \ref{NRGEvsT}  characterize an instability that is not predicted by the dispersion theory.  Shown are electric field energy histories  for a cold, drifting plasma for a sequence of values of $B$. 
In Fig. \ref{NRGEvsT} (a), $B$ is varied while $C=0.08$ is fixed. That is, $\Delta x$ is varied but $U=0.16$ and $\Delta T/\Delta x=0.5$ are held fixed.  The  values of $B$ are:   
\begin{equation}
B=[0.16, 0.32, 0.64, 1.28, 2.56, 5.12, 10.24, 20.48] .
\label{Bsequence}
\end{equation}
 All the histories unfold similarly in 3 phases:  In the first phase, there is exponential growth of the field energy.  The growth rate is, as advertised, the same for all the values of $B$.  In the second phase some process, most likely particle trapping, ends growth.  Finally, in the third phase the field energy is in steady state. The histories are ordered from top to bottom in increasing $B$.  Each curve corresponds to a simulation with double the value of $B$ and 1/4 the value of the field energy relative to the total energy for the curve above it.  Since the total energy is constant,
the value of the energy at saturation varies as $1/B^2$. 
In Fig. \ref{NRGEvsT} (b), we compare the field energy at saturation with the total energy error  for $C=0.8$ and $B$ from Eq. \ref{Bsequence}.  Both the field energy, $\Box$'s, and energy error, $\diamond$'s, vary as $1/B^2$ .  In fact, except for the point corresponding to $B=0.16$, which is unstable to the FGI, the ratio of the field energy to the total energy error is nearly constant.  For example, the ratio of the field energy to the energy error for $B=2.56$ is $0.86$, and for $B=20.48$ is $0.87$.  Plasma heating makes a small contribution to the error, and it also varies as $1/B^2$.

\begin{figure}

\includegraphics[width=100mm]{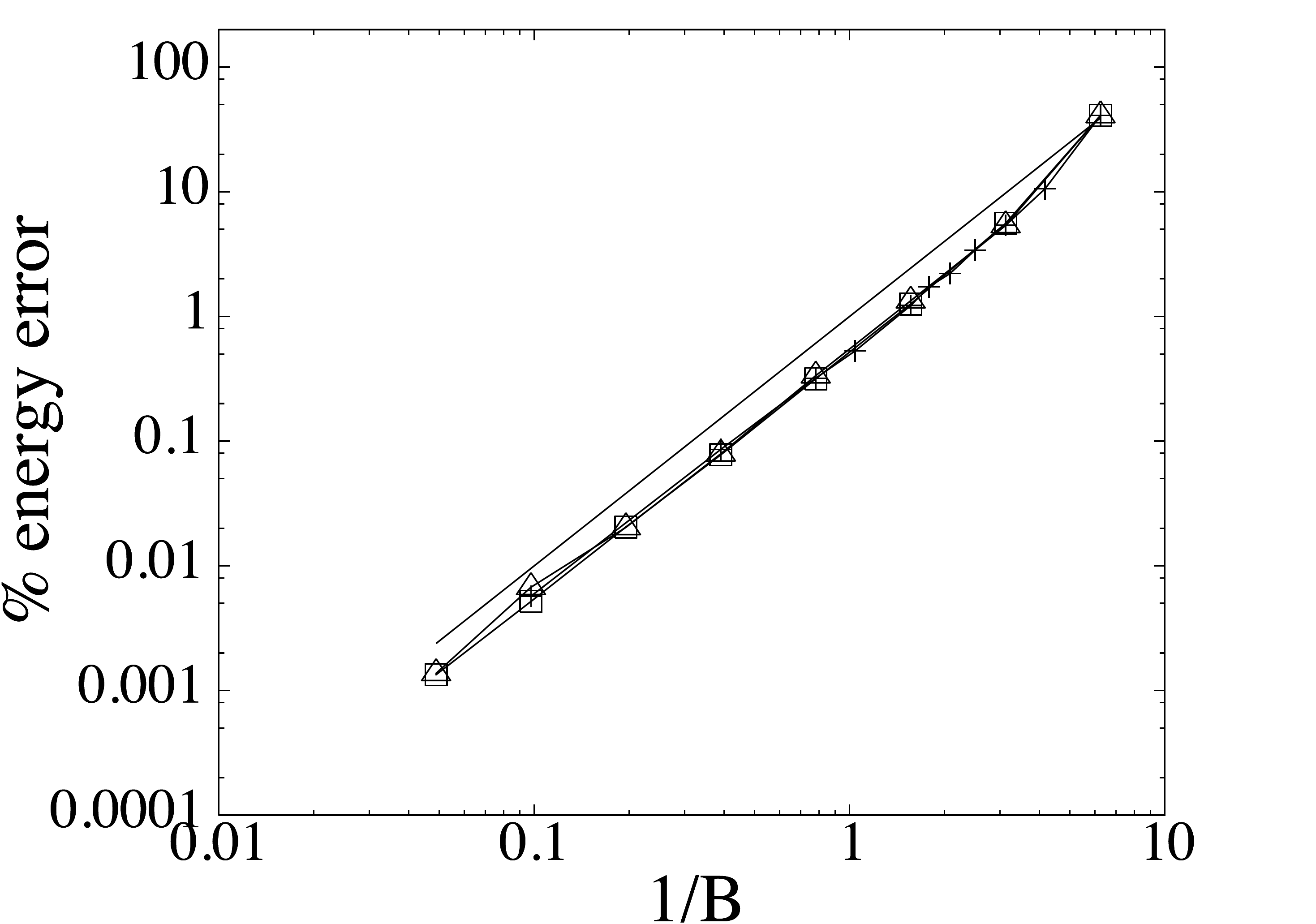}
 \caption{  $\epsilon$ from CIC simulations of a cold, drifting plasma scales as $1/B^2$ no matter how the numerical parameters are varied.  $B$ is varied:   by varying $\Delta x$ and $\Delta T$ with $\Delta T/ \Delta x$ constant ($\Box$'s);    by varying $\Delta x$ with constant $\Delta T$ ($\triangle$'s);  by varying $U$ with constant $\Delta x$ and $\Delta T$ ($\ast$'s,).  The straight line $\epsilon=1/B ^2$ is included for reference.   } 
\label{NRGconverge}
\end{figure}

In Fig \ref{NRGconverge} the results show that the energy error converges to $0$ as $1/B^2 \to 0$.   Three sequences of simulations are compared with each other and with the straight line $1/B^2$. In the first sequence ($\Box$'s), $C=0.08$ and $B $ is from Eq. \ref{Bsequence}.  As in Fig. \ref{NRGEvsT} , $U=0.16$ and $\Delta T/\Delta x=0.5$.   The error varies as $1/B^2$.  In the second sequence ($\triangle$'s),  $B$ is again given by Eq. \ref{Bsequence} but $C=B \Delta T$.  In this sequence, $U=0.16$ as before but $\Delta T=0.5$ and $\Delta x$ is varied. The error still varies as $1/B^2$.   In the third sequence ($\ast$'s) $\Delta x=c/\omega_{pe}$, $\Delta T=0.5$ and $U$ is varied. In this sequence, the absolute error is constant but the relative error varies as $1/B^2$ because the total energy varies as $B^2$.   

The scaling of errors in energy conservation is explained by the re-centering of the electric field,  Eq. \ref{E_c}. Namely, only error terms that come from replacing $E_v$ by  $E_c= 1/2 ((E_{v+1}+E_v)$ in the CIC particle momentum equation contribute.  Other error terms that depend on $\Delta T$, are absent  because they are proportional to the pressure, Eq \ref{pressure}, which is vanishingly small for a cold beam.

\subsection{Warm, stationary plasma}

\begin{figure}
\begin{tabular}{cc}
(a)  &  (b)  \\
\includegraphics[width=60mm]{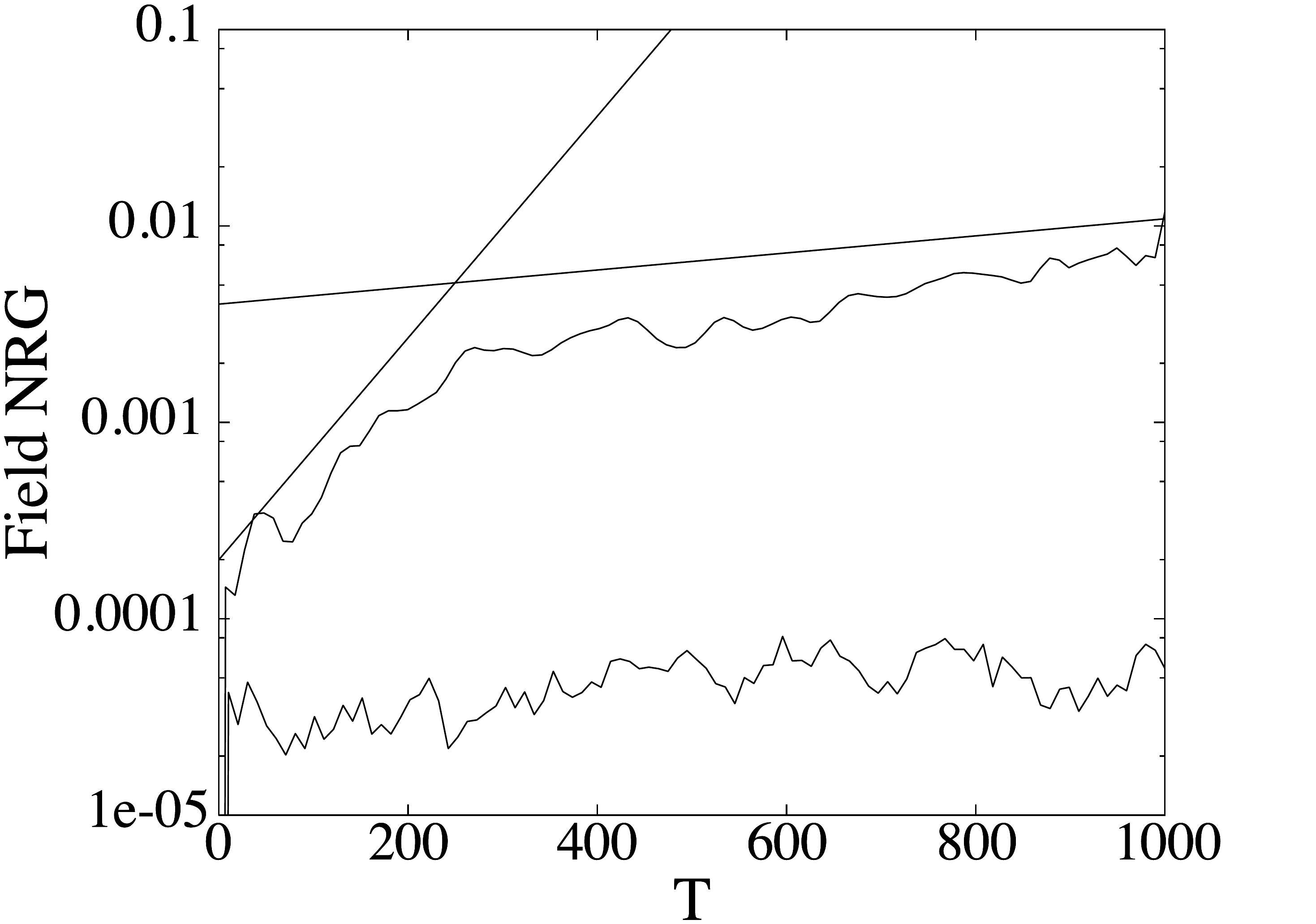}   &
\includegraphics[width=60mm]{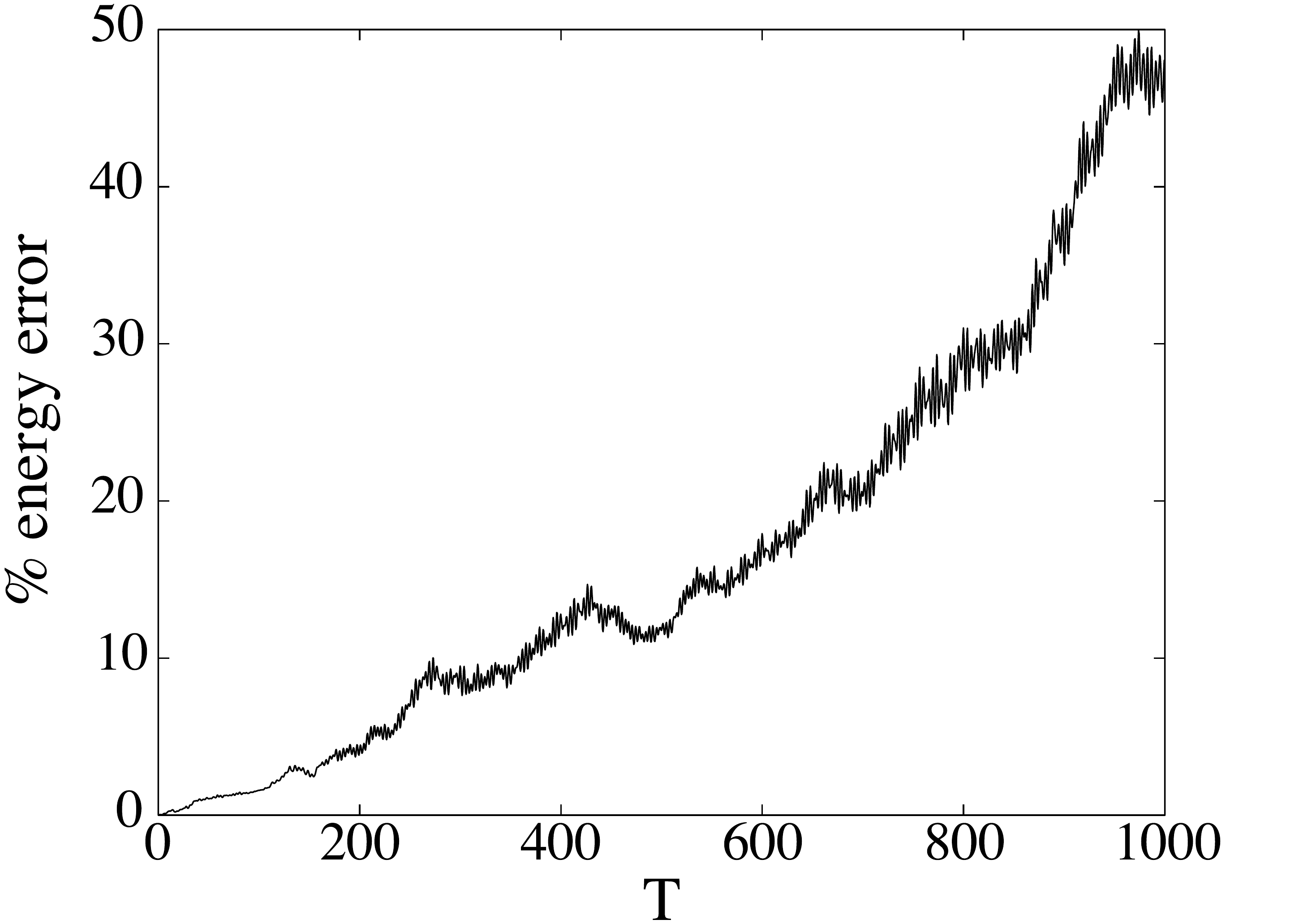}
\end{tabular}
\caption{ $\mathcal{F}$ in a stationary, Maxwellian plasma is shown in (a)  for CIC (upper curve) and for CELESTE (lower curve). The straight lines are the FGI growth rates for modes 2 and 4.   $\epsilon$ for CIC is shown in (b).}
\label{FGI_mode4}
\end{figure}

We repeat portions of Okuda's  study in one dimension \cite{okuda1972}, and compare the energy and momentum errors for CIC with those for CELESTE.  (This is the only case we have found for the growth of the FGI in a Maxwellian plasma.)
A warm, stationary plasma is simulated on a  periodic domain with length $L=8 c/\omega_{pe}$ with fixed ions.  The initial particle velocities are sampled from a Maxwellian distribution.   The numerical parameters are $D=0.10$, $B=0.0$, $C=0.0125$, and $N_D=30$.
Okuda's simulations with CIC end at $T=40$ \cite{okuda1972}.  Ours end at $T=1000$, much longer but still a short time compared with current practice.
In Figure \ref{FGI_mode4}(a), the growth of the electric field energy with CIC (upper curve) is compared with the functions $f_1=0.0002 \times  exp(0.013 T)$ corresponding to mode 2 with $k \Delta x=\pi/2$ and $f_2=0.004 \times exp(0.001 T)$ corresponding to mode 4 with $k \Delta x=\pi/4$ \cite{langdon1970}, and with the much smaller and more slowly growing CELESTE electric field energy  (lower curve).  The faster growing mode saturates at $T \approx 250$;  the more slowly growing mode has not saturated by $T=1000$.    As Okuda remarked \cite{okuda1972}, even when a numerical instability grows slowly, it can result in significant heating over longer periods, Figure \ref{FGI_mode4}(b). With CELESTE, there is no heating in this case. 

\begin{figure}
\includegraphics[width=100mm]{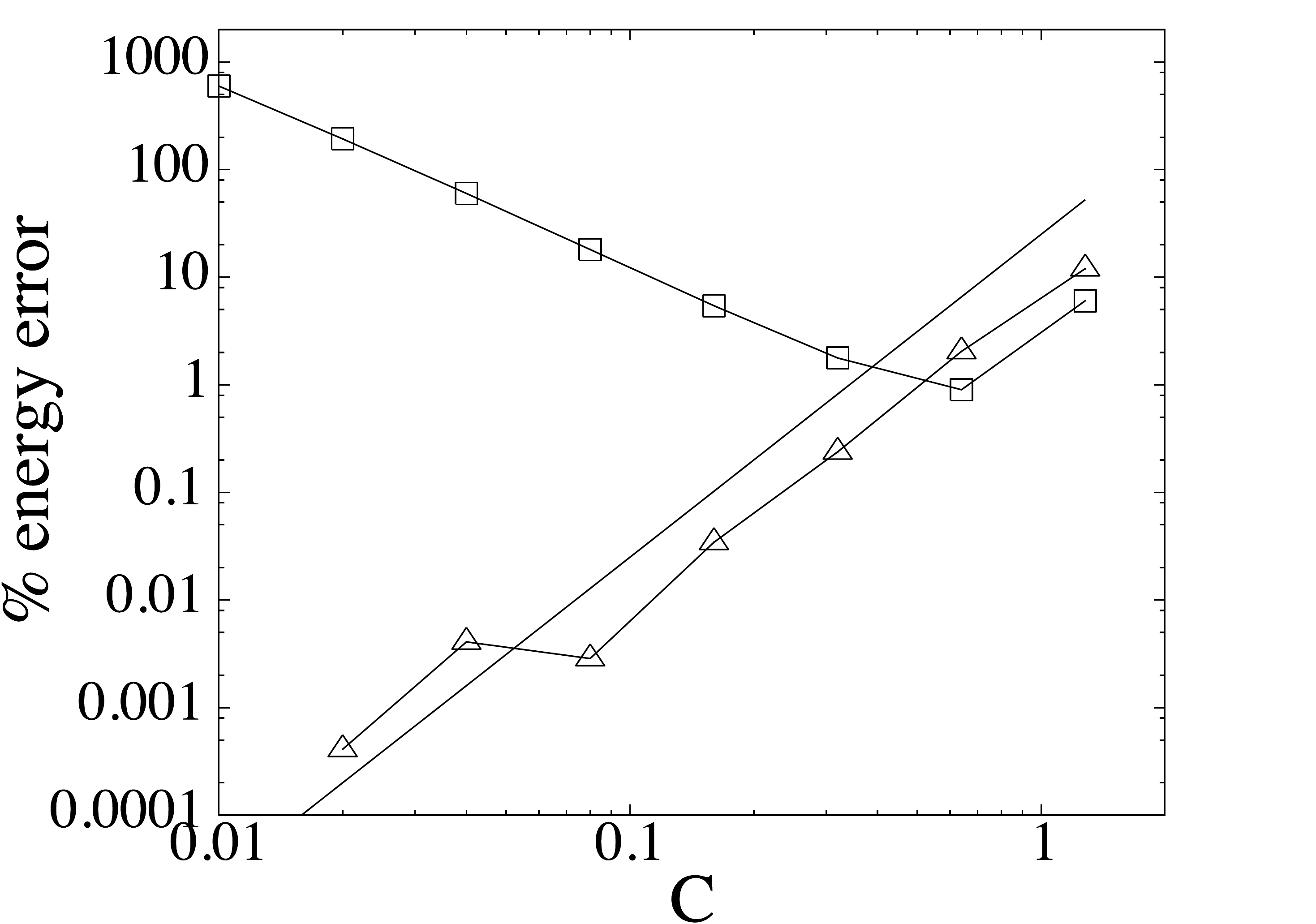} 
\caption{  
$\epsilon$ for CELESTE ($\bigtriangleup$) and CIC ($\Box$) in simulations of a warm, stationary plasma  are compared for $C/D=\Delta T=0.125$. For $C>1$, the CELESTE $\epsilon$ ( $\bigtriangleup$)  is twice that for  CIC ($\Box$).  The CIC $\epsilon$ varies approximately as $1/C^2$ for $C<0.64$, and as $C^3$ for $C>0.64$.  The CELESTE error varies as $C^3$ for all  $C$.  }
\label{PICLN_Cscaling}
\end{figure}

When $D=0.16$,  a case that  Hockney studied \cite{hockney1971}, the FGI  growth rate from theory is much smaller than with $D=0.1$.  With  $D=0.16$, $B=0$, $\Delta T=0.5$, $C=0.08$, and $N_D=20$ the FGI seems to be much weaker.   \begin{figure}
\begin{tabular}{cc}

(a)  &  (b)  \\
\includegraphics[width=60mm]{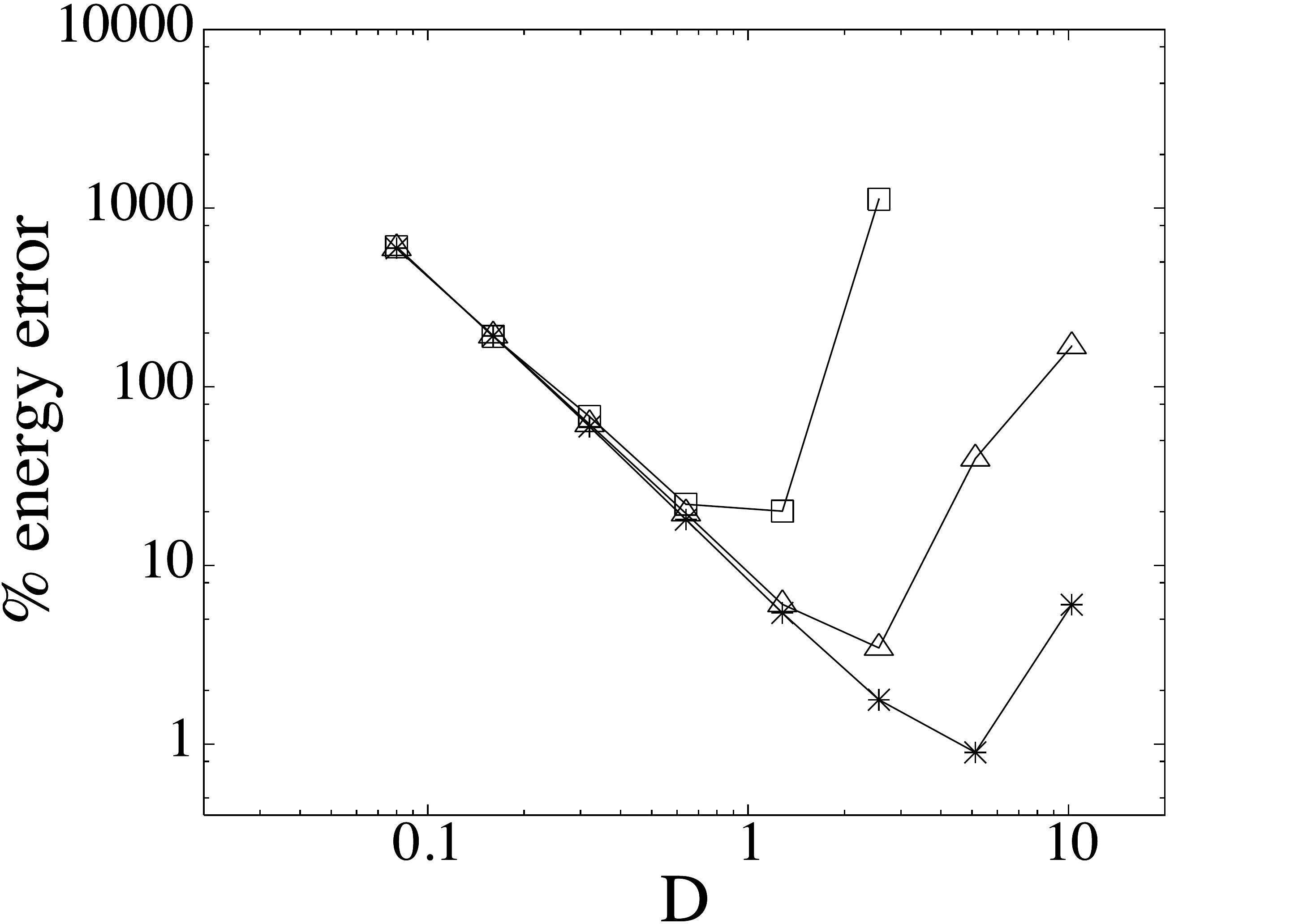}   &
\includegraphics[width=60mm]{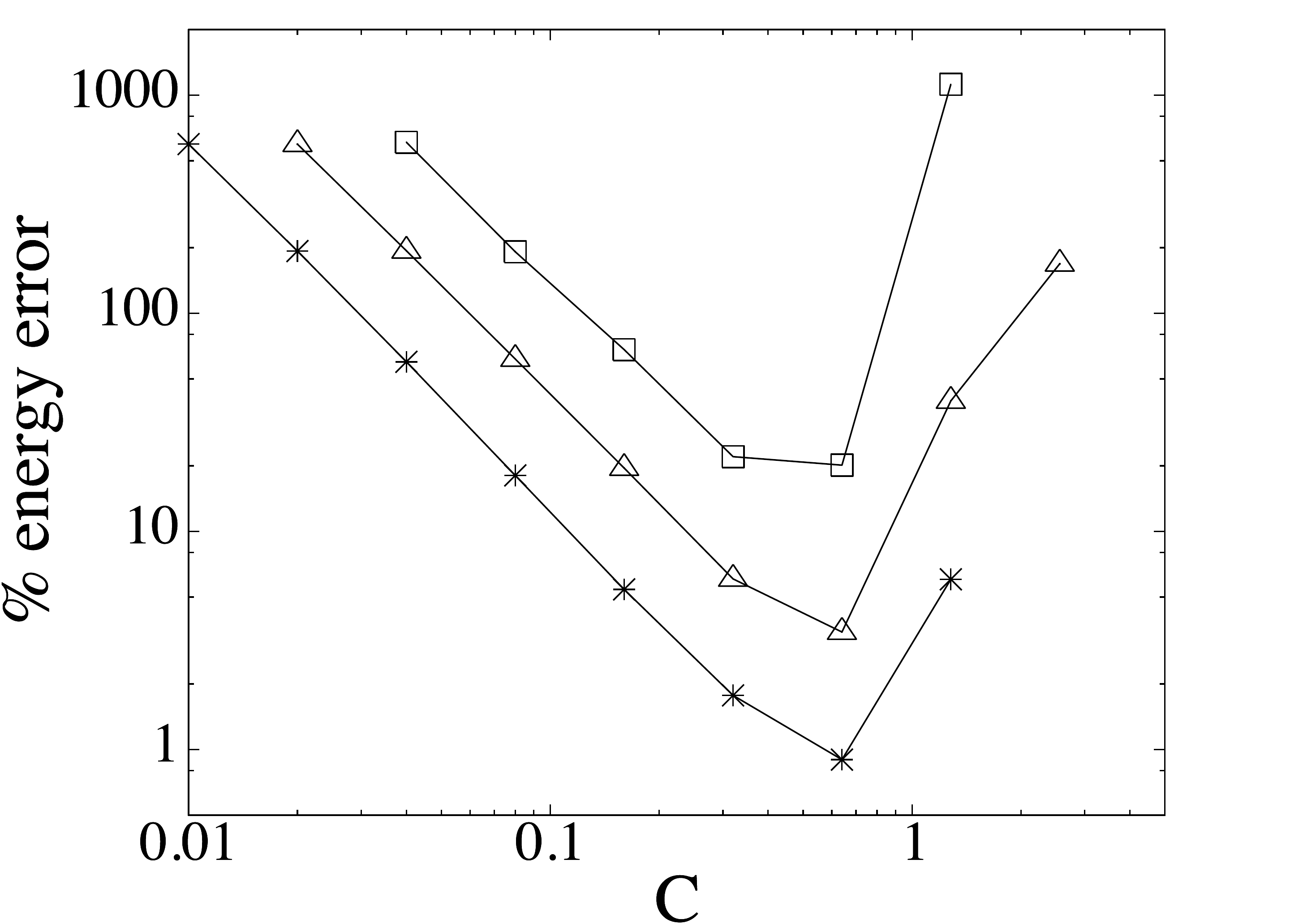} 
\end{tabular}
\caption{In (a) with CIC), $\epsilon$  depends on $\Delta x$  but not $\Delta T$ for values of $D$ corresponding to the minimum error and below.   There $\epsilon$ scales as $1/\Delta x^2$.  The 3 time steps for which data is plotted are  $ \Delta T = 0.5$  ($\Box$), 0.25 ( $\bigtriangleup$), and 0.125 ($\ast$).In (b), data from (a), are replotted against C. $\epsilon$ is smallest when $C \approx 0.64$. Hockney \cite{hockney1971} notes that for $C=1$, $60 \%$ of the particles cross cell boundaries each time step.  }
\label{CICNRGD}
\end{figure}
In Figure \ref{PICLN_Cscaling},   
the values of $\epsilon$ computed by CIC and by CELESTE are compared for $C/D=\Delta T =0.125$.  For $C>0.64$, CIC error varies as $C^3$.  For all $C$, CELESTE error varies as $C^3$.  It is often stated that CELESTE, and energy-conserving methods in general, conserve energy exactly only when $\Delta T \to 0$, but it should be noted the errors are small compared with those for CIC when $C < 0.3$.

\begin{figure}
\includegraphics[width=100mm]{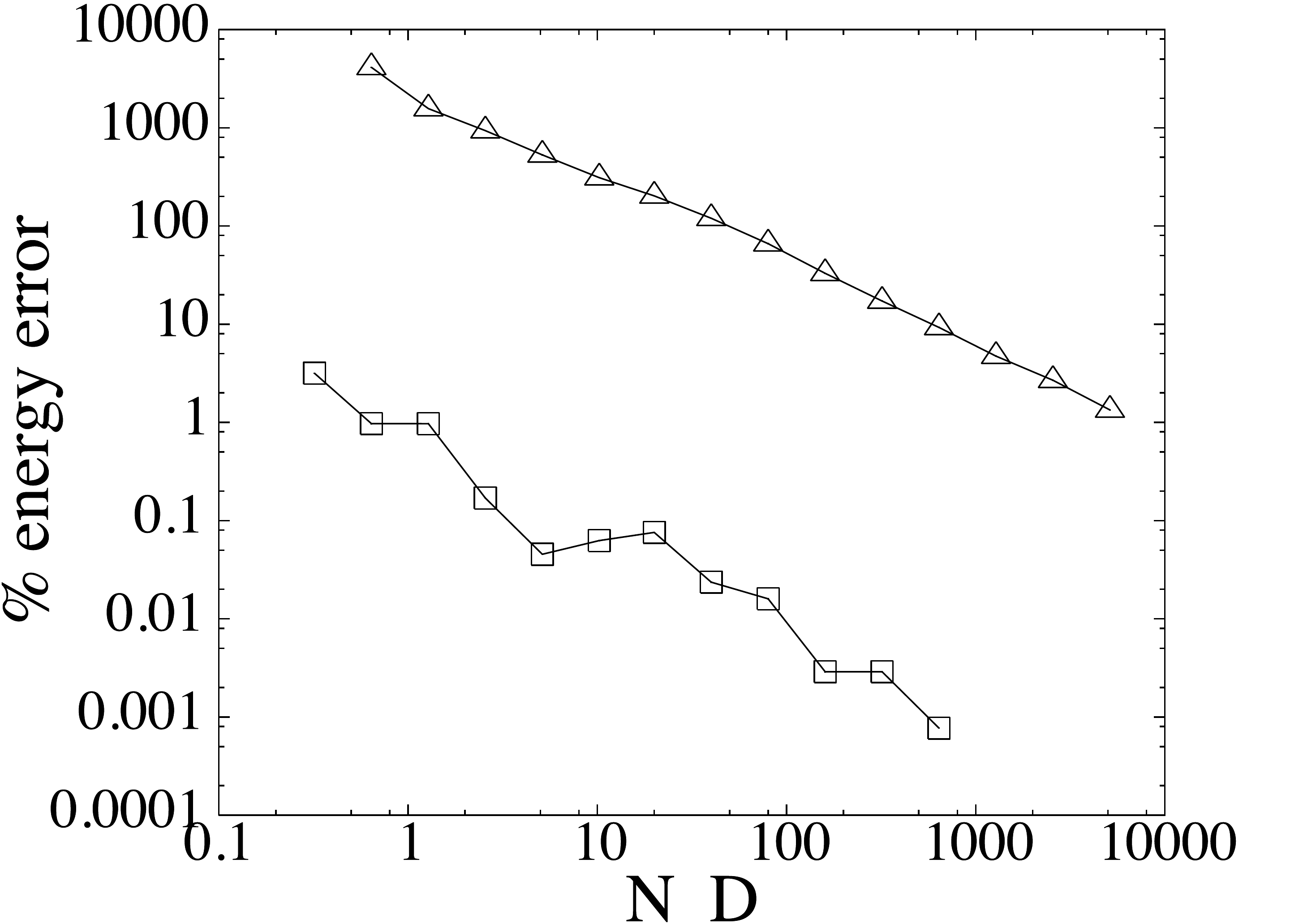} 
\caption{ Numerical heating is much smaller for CELESTE( $\Box$) than for CIC, points labeled ($\ast$).  Heating scales for both as$1/N_D$.  $D=0.16$, $\Delta T= 0.5$, $C=0.08$.  }
\label{N_pPICL}
\end{figure}

The scaling of the  energy error, or heating in this case,  with $N_D$  is shown in Figure \ref{N_pPICL} for both CIC and CELESTE. For all values of $N_D$,  $D=0.16$,  $\Delta T=0.5 $ and $C=0.08$.    For CIC in Figure \ref{N_pPICL}, the heating decreases from $5000 \%$ of the initial energy with $N_D=1$ to $1 \%$ with $N_D=5120$, and varies linearly with  $1/N_D$ in between.   
For CELESTE, the heating also varies as $1/N_D$  (lower curve in Figure \ref{N_pPICL} with open squares), but for all values of $N_D$, the heating for CELESTE is several thousand times smaller than for CIC.   

 For an equilibrium plasma, the momentum errors are discussed in \cite{hockney1971}[pp252-253].  A random walk argument is used to show that the secular drift of the momentum should not be a serous problem for energy-conserving methods (like CELESTE)
if $C < 1$ and $N_D >>1$.  Experiments with CELESTE give random errors in the average drift that are less than $1 \%$ of the thermal speed..

\subsection{Warm drifting plasma}

We now consider a warm, drifting plasma  with $D=B$.  This is a more typical plasma condition than the cold beam in Okuda's problem \cite{okuda1972}, and the FGI does not seem to occur.
Nevertheless, there are numerical errors that cause a CIC plasma  to heat and a CELESTE plasma to lose momentum.

In Hockney \cite{hockney1971},  computational results verify the the dependence of collisions and heating on the number of particles.  The collision and heating rates scale inversely as the number of computational particles in a Debye length.

\begin{figure}
\begin{tabular}{cc}
(a) & (b)  \\
\includegraphics[width=65mm]{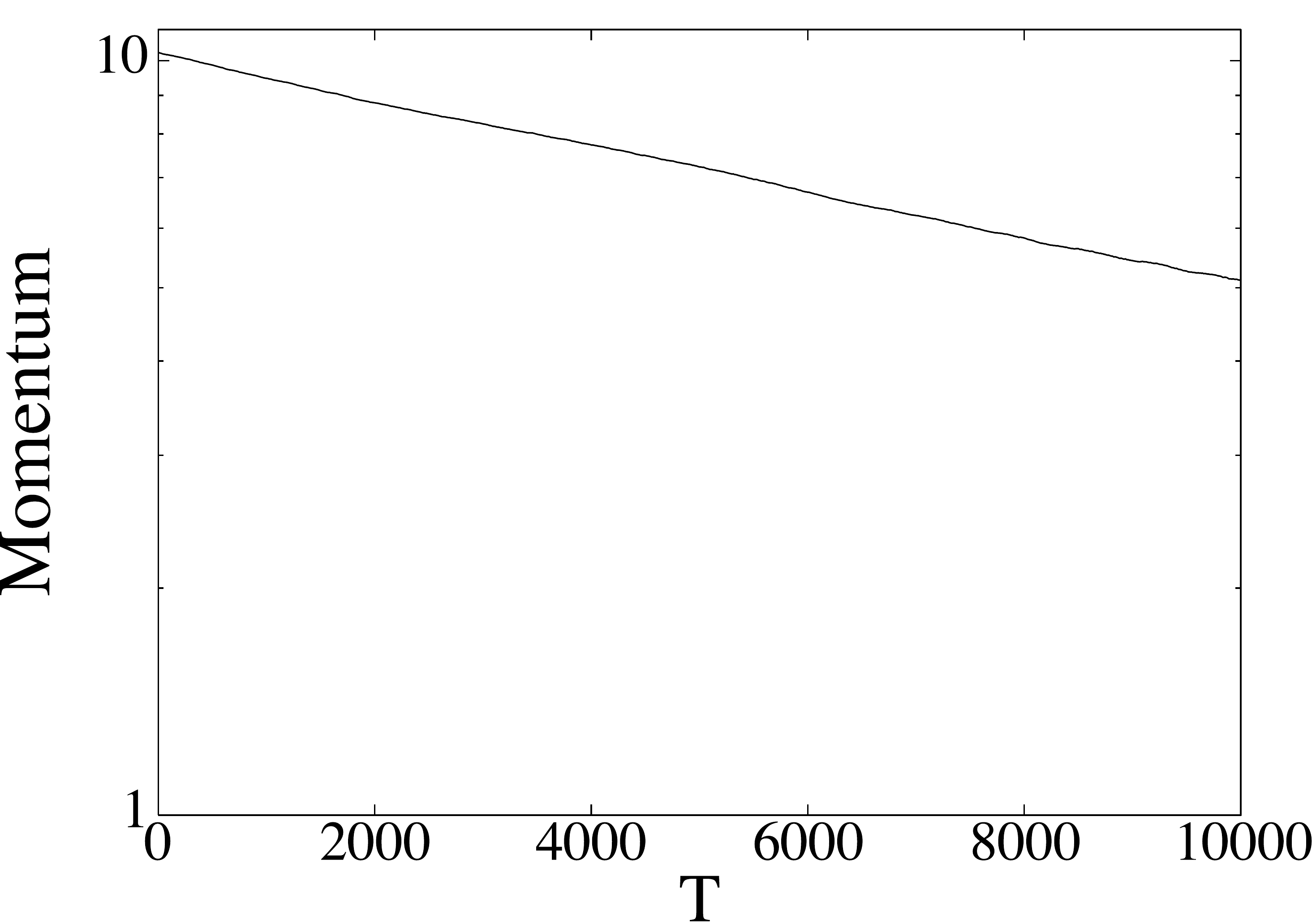} &
\includegraphics[width=65mm]{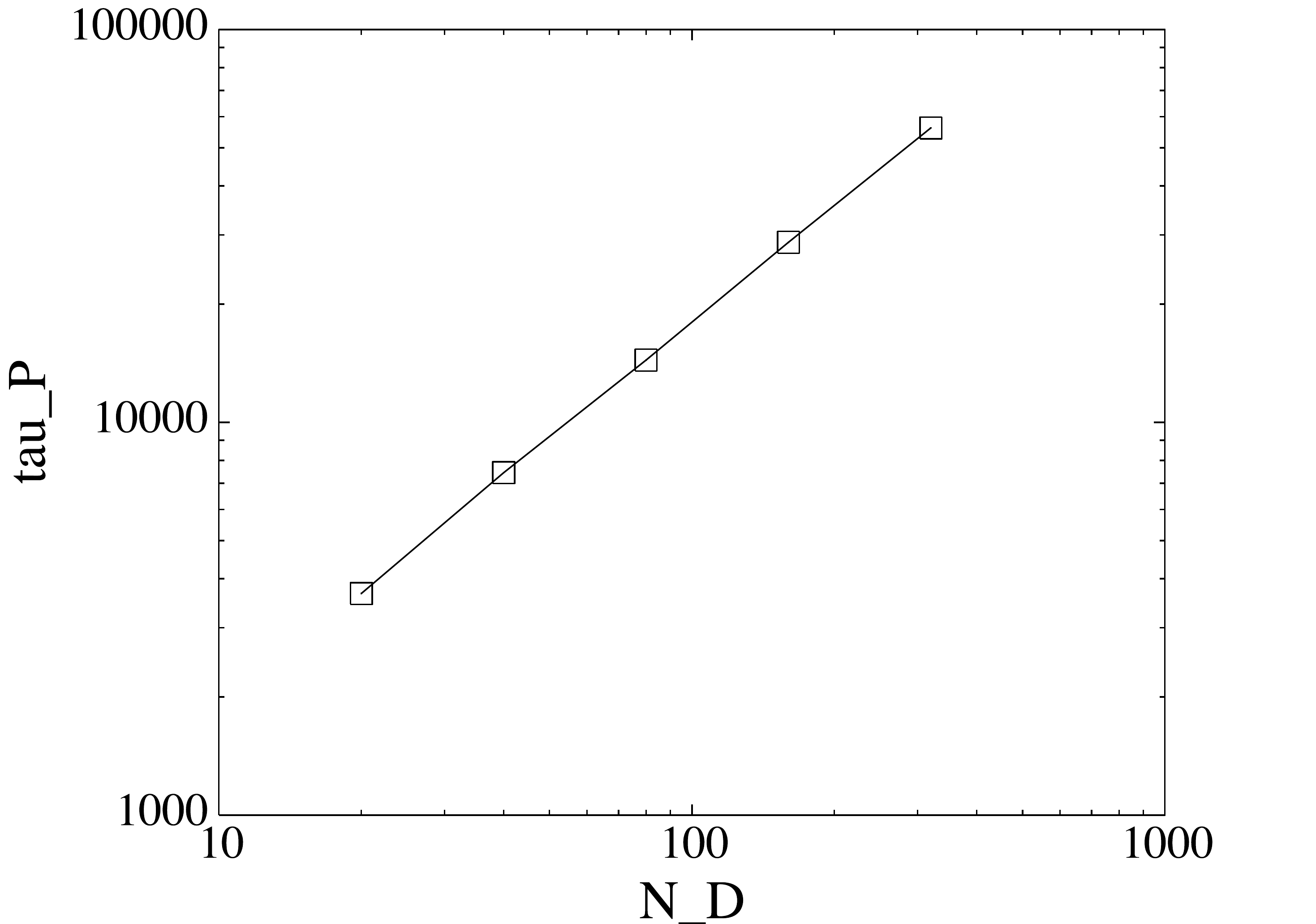}
\end{tabular}
\caption{The CELESTE momentum decays exponentially with time (a).   The CELESTE momentum decay  time, $\tau_P$, increases as $N_D$ (b) }
\label{TauPvsND}
\end{figure}

For CELESTE, the momentum decay slowly in Figure \ref{TauPvsND}(a) compared with trapping for a cold beam Figure \ref{Momentum}. The warm beam momentum decays exponentially, due to a mechanism that operates continuously. The data in Fig. \ref{TauPvsND} (b) in which the momentum decay time scales as $N_D$ suggests the momentum decay is caused by charge fluctuations.

\begin{figure}
\begin{tabular}{cc}
(a) & (b)  \\
\includegraphics[width=65mm]{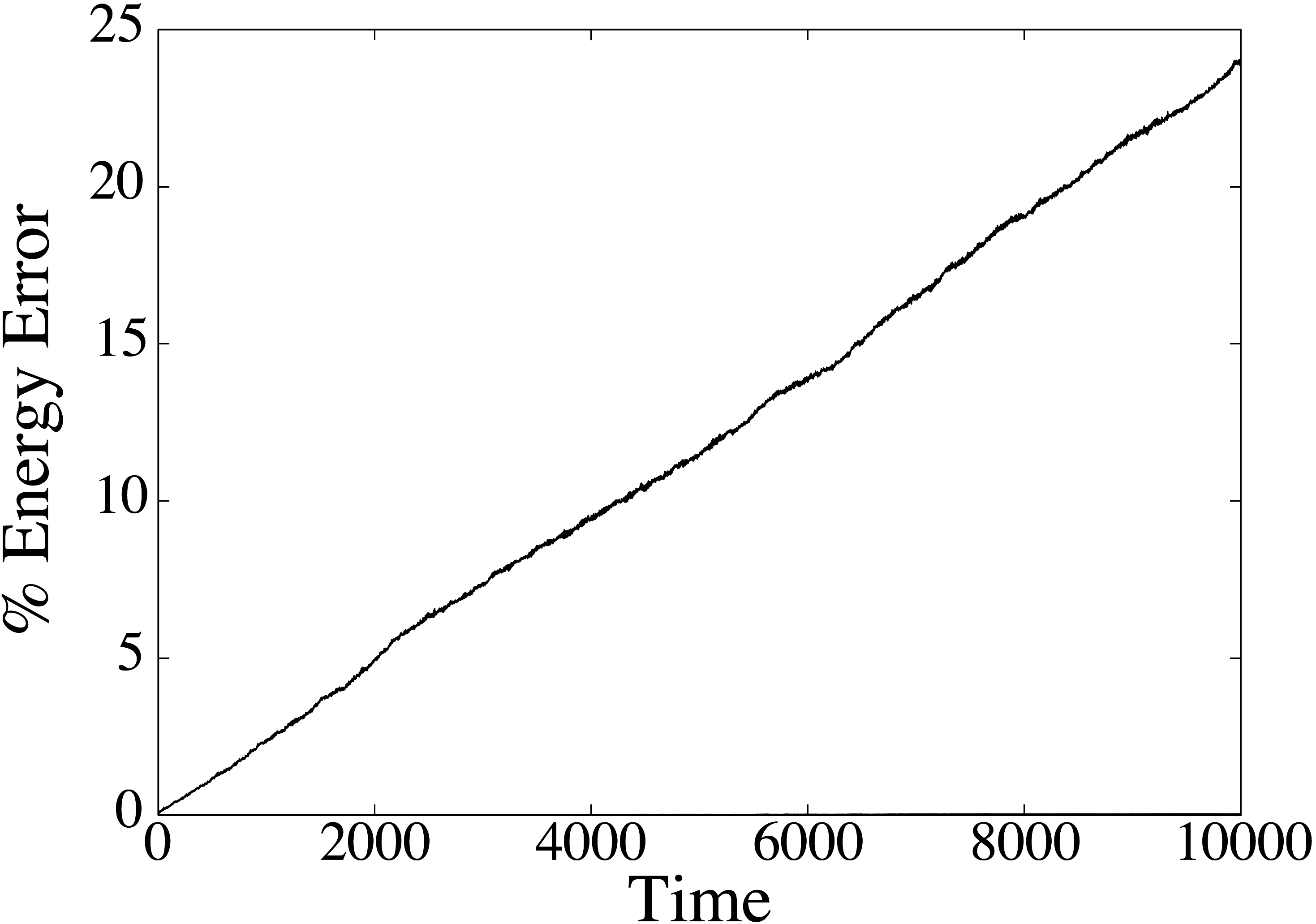}   &
\includegraphics[width=65mm]{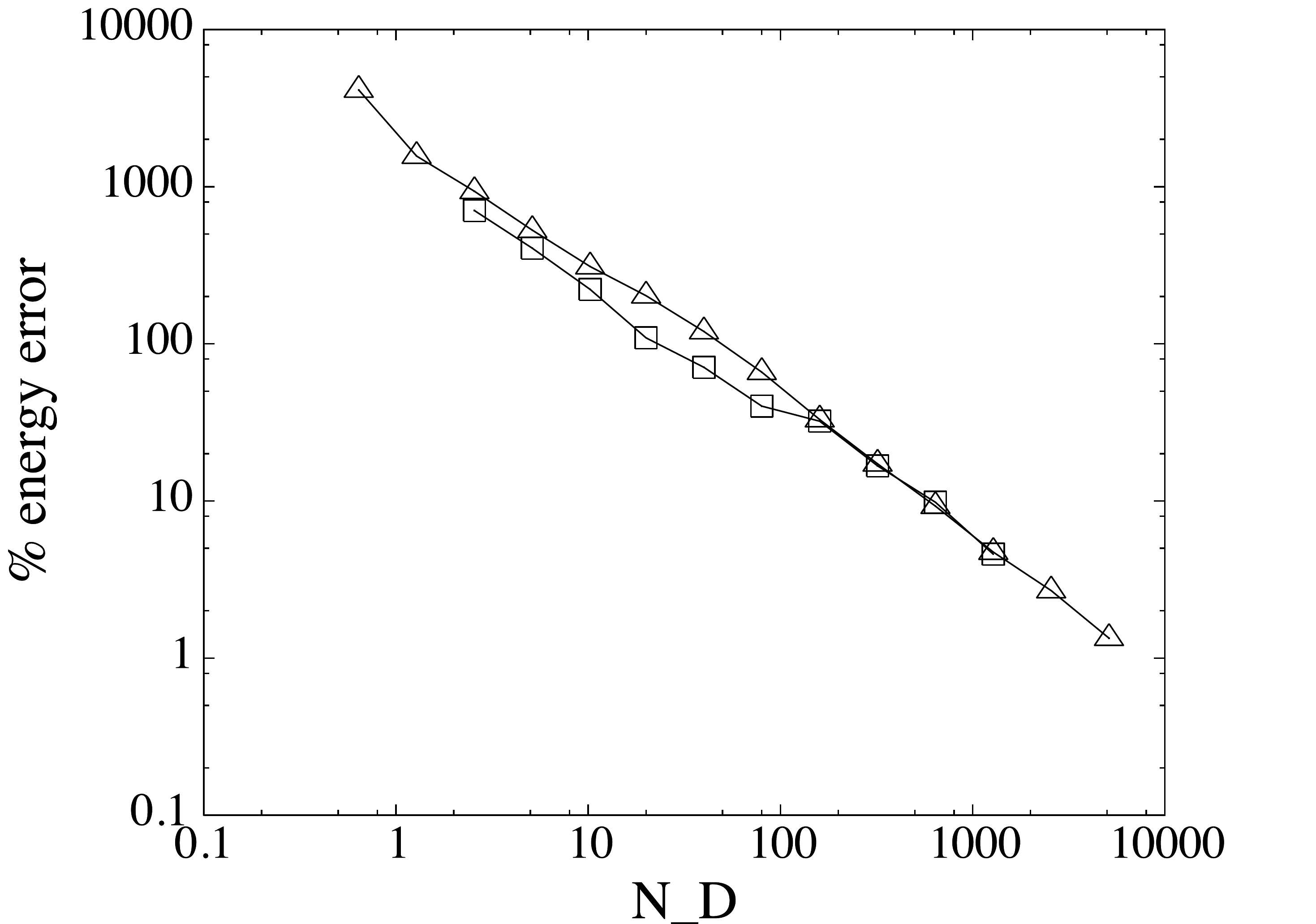}
\end{tabular}
\caption{Due to numerical error, $\epsilon$ for a CIC simulation of a warm plasma increases linearly with time (a). The CIC heating rate in (b) varies as $1/N_D$  for both drifting  ($\Box$'s) and stationary ($\bigtriangleup$'s) plasmas.  (The stationary data is reproduced from Figure \ref{N_pPICL}.)}.  
\label{EpsvsND}
\end{figure}

 With CIC, the energy error increases linearly in time, Fig. \ref{EpsvsND}(a), which agrees with earlier results in \cite{hockney1971,cohen1989}.  The error scales as  $1/N_D$, Fig. \ref{EpsvsND}(b)
 for $5 \leq N \leq 5000$.  With a warm beam, the momentum and energy errors decrease as the number of particles per Debye sphere increases.

Fig. \ref{EpsvsND}(b) compares the variation in energy error in CIC simulations with $N_D$ for a warm, drifting plasma with $U_0=v_{te}=0.16$, and the stationary plasma with $U_0=0$ and $v_{te}=0.16$. The energy errors for both are computed relative to the thermal energy.  The drift energy is subtracted.  Thus, the absolute errors are comparable in size and in their decrease as $N_D^{-1}$. Consider the energy error that might be produced by a finite grid instability.  The FGI error can contribute to energy error in drifting but not stationary plasma which might account for the differences in value between moving and stationary results except that  an FGI-caused error is independent of $N_D$.

\begin{figure}
\includegraphics[width=100mm]{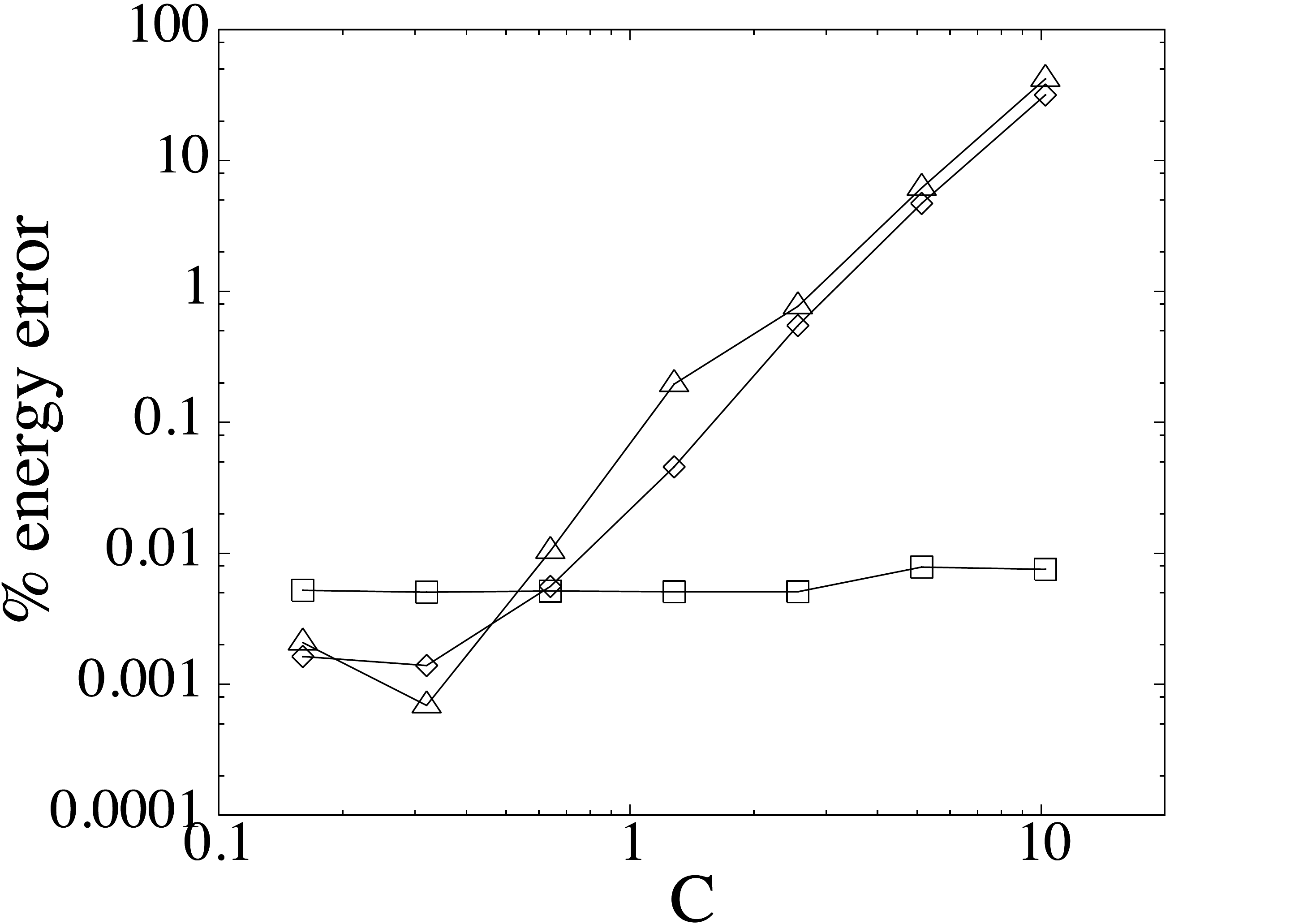}
\caption{  $\epsilon$ from CIC simulations of a cold, drifting plasma ($\Box$'s), a warm stationary plasma ($\bigtriangleup$'s), and a warm, drifting plasma ($\diamond$'s) are shown. The error for the warm plasmas, stationary or drifting, varies as $C^3$.  The error for the cold plasma is nearly independent of $C$.   } 
\label{NRGHvsC}
\end{figure}

In Fig. \ref{NRGHvsC}, the CIC energy errors for  cold drifting,  warm drifting, and cold stationary  plasmas are shown.   All of the simulations have $N_D=500$,  and $D=10.24$.  $\Delta T$ varies so that  $0.16 \leq C \leq 10.24$.  For a warm plasma with $v_{te}=0.16$ that is either drifting with $U=0.16$ ($\diamond$'s) or stationary with $U=0$  ($\bigtriangleup$'s).  For a cold drifting plasma with $v_{te}=0$ and $U=0.16$,   ($\Box$'s), the energy error is independent of $C$,  unlike the corresponding plots for hot plasmas, both stationary and drifting, where  $\epsilon$  varies as $C^3$.  The difference is caused by the absence of a higher moment contributions to the current in the cold drifting beam, Section \ref{NRGerrors}.  With no thermal velocity, all higher order terms in the moment equation expansion, Eq. \ref{Qcontinuity}, of the continuity equation are zero, and there is no truncation error.  In addition, $\epsilon$ is larger at small values of $C$ for a cold plasma than a warm plasmas.  This is likely due to the contribution of an FGI.  The CIC cold plasma is unstable to an FGI (Section \ref{aliasing}),  $\epsilon \approx 0.01 \%$ for $D=10.24$ (Figure \ref{NRGEvsT} b), and saturation of the FGI is independent of $\Delta T$.

\section{Conclusions}

The physics model in Section \ref{Model} conserves both momentum and energy if Gauss' law is satisfied everywhere.  The PIC algorithms in Section \ref{Algorithms} satisfy Gauss' law either at cell centers or at cell vertices.  For momentum-conserving algorithms, such as CIC  with cell-centered differencing, Gauss' law is satisfied at cell centers and momentum is conserved to round-off.  For energy-conserving algorithms, such as CELESTE or LEWIS, the derivation of the equations from Hamilton's principle results in a staggered mesh scheme in which Gauss' law is satisfied at cell centers by the solution of the potential equation but not at cell vertices where momentum is computed. 

When Gauss' law is satisfied, the total force acting on the particles can be expressed as the divergence of a Maxwell stress tensor, Eq. \ref{EMaxwell}.  To reproduce this result with difference equations requires that summation by parts of $(\nabla \cdot \E) \E$ result in a collapsing sum, one that depends only on the boundary conditions.  A problem with some similarities occurs in magnetohydrodynamics, where it is required that $\nabla \cdot \B=0$ for conservation of momentum.  There one can choose to express the force exerted by the magnetic field on the plasma as $\bJ \times \B$, which does not conserve momentum, or by a Maxwell stress tensor, in which case there may be non-physical contributions from $\nabla \cdot \B \ne 0$ \cite{brackbill1980}.  

Momentum conservation or non-conservation is not altered by the introduction of implicit differencing in time.   In recent work on implicit, energy-conserving methods, for example  \cite{markidis2011,chen2011},  errors in momentum conservation are identified in the numerical examples.  However, implicit methods allow large time steps compared with the plasma frequency, the contributions of fluctuations in net charge vary inversely with the time step \cite{mason1981}, and so errors in momentum conservation, which depend on errors in Gauss' law, don't grow with the time step. 

Probably more plasma heating is attributed to the FGI than should be. The FGI seems to cause larger energy errors for cold drifting plasmas, Figure \ref{NRGCIC}, than for warm plasmas, either drifting or stationary.  For example, in a warm plasma with $D=0.16$,  Figure \ref{N_pPICL}, heating is observed to vary as $1/N_{D}$  for both CIC and CELESTE, and the FGI is independent of $N_D$. If the spatial resolution is sufficiently coarse, for example with $D = 0.1$, a weak FGI with CIC, (and none with CELESTE), causes significant heating with CIC (but not with CELESTE), Figure \ref{FGI_mode4}b.    

The FGI causes momentum loss in LEWIS and CELESTE simulations of a cold, drifting plasma, Figure \ref{Momentum}, by causing electron trapping.  In general, however, any mechanism physical or numerical that causes spatial variations in the net charge will cause errors in momentum conservation unless $D$ is sufficiently large, as it is in the simulation of the Buneman instability, Figure \ref{ENRG}.  If one argues that the stability boundary for the FGI is useful guide, then the spatial resolution should satisfy $D>0.2$ with CELESTE, or $D>0.4$ with LEWIS.  For improved energy conservation, spatial resolution with CIC should satisfy a similar but more restrictive condition $D \approx 1$, Figure \ref {CICNRGD}.

The energy analysis, Section \ref{NRGerrors}, gives for CIC energy error scaling as $C^2$, Eq. \ref{averaging} and for CELESTE as $C^3$, Eq. \ref{no-avg}.  The numerical results, Figures \ref{PICLN_Cscaling} and \ref{NRGHvsC}, seem to show that CIC and CELESTE energy errors both scale as $C^3$.    
This appears to disagree with \cite{vu1992}, where numerical results suggest that quadratic charge assignment, as in CELESTE,  gives smaller energy errors than linear assignment, as in CIC, and analysis shows that the greater continuity of a quadratic function reduces errors.  The results in \cite{vu1992} are for a different problem and no scaling is given, so a direct comparison is not possible.

Two obvious questions are left unresolved by this work.  There is no obvious explanation for the disagreement between theory and CIC results for the FGI, Figure \ref{CICGamma}(a) and \cite{birdsall1980}.   Why an FGI-like instability should develop for $k \Delta x=\pi$ where theory says the growth rate should be zero suggests there is something essential left out of the dispersion theory, but not what that might be.  

The big unresolved question is how one might formulate a simulation algorithm that conserves both momentum and energy.  The model, Section \ref{Model}, is clearly conservative, and  the analysis in Sections \ref{momentum} and \ref{NRGerrors} makes clear what properties the finite difference equations must have to preserve the constants of the motion.  However, it is not just a question of finite differences.  The assignment of particle properties to the grid constrains the difference equations, and achieving conservation is more complex as a result.  It's a fascinating puzzle.

\section*{Appendix:  Some Useful Properties of b-Splines}

\label{appendix}

 The PIC formulation
 uses the special properties of the b-spline \cite{deboor1978}.  Their properties are described in many references, among them \cite{haugbolle2013}.  Here we list those properties that are relevant to a comparison of energy-conserving and momentum-conserving PIC formulations.
 
The PIC equations in one dimension are solved on the domain $x\in[0,L]$, which is divided into $N$ cells with width $\Delta x$.  Cell vertices are labeled $x_v: v \in [1,N+1]$ and cell centers $x_c=1/2(x_{v+1}+x_v)$.

For any $l$, $\Snp1$ is equal to the convolution of $\Sn$ with $\Sngp$,
\begin{equation}
\Snp1=\Sn \ast \Sngp,
\label{convolution}
\end{equation} 
where $\Sngp$ is a characteristic function with support $\Delta x$,
\begin{equation}
\Sngp(x; \Delta x) = \left \{ \begin{array} {ccc} 1/\Delta x & if & x \in [\Delta x/2,\Delta x/2],  \\
      0  &if&x \notin       [\Delta x/2,\Delta x/2].  \end{array}   \right.
\end{equation}
If the support is not specified, it is assumed to be equal to $\Delta x$.

Analytic differentiation of $\Snp1$ yields a centered difference equation,
\begin{equation}
\frac{\partial \Snp1}{\partial x}=\frac{\Sn(x+\Delta x/2)- \Sn(x-\Delta x/2 )}{\Delta x}
\end{equation}
Finally, we note that $S$ is normalized so that for all $l \geq 0$,
\begin{eqnarray}
1 &=& \int \Sn(x) \dx,     \nonumber \\
1/\Delta x & = & \sum_c \Sn(x-x_c)= \sum_v \Sn(x-x_v).  \nonumber
\end{eqnarray}

The Fourier transform of $\Sngp$ is \cite{birdsall1985},
\begin{equation}
\mathcal{S}_k^{(0)}= \frac{sin (k \Delta x/2)}{k\Delta x/2}.
\end{equation}
Using the Eq. \ref{convolution}, 
\begin{equation}
\mathcal{S}_k^{(l)}=\left ( \frac{sin (k \Delta x/2)}{k\Delta x/2} \right)^2.
\end{equation}

\section*{Acknowledgements}
I am grateful for many useful discussions with Gianni Lapenta, Gabor Toth and Sergey Gimelshein.

\pagebreak


\begin{thebibliography}{99}

\bibitem{dawson1962}
J. Dawson, One-dimensional plasma model, Phys. Fluids {\bf 5} 1962 445-459.

\bibitem{birdsall1985}
C. K. Birdsall and A. B. Langdon, Plasma Physics via Computer Simulation, Adam Hilger, Philadelphia, 1991.

\bibitem{hockney1988}
R. W. Hockney and J. W. Eastwood, Computer Simulation Using Particles, Institute of Physics Publishing, Philadelphia, 1988.

\bibitem{grigoryev2002}
Yu. N. Grigoryev, V. A. Vshivkov, and M. P. Fedoruk, Numerical 'Particle-in-Cell' Methods, VSP, Utrecht, 2002.

\bibitem{dawson1995}
J. M. Dawson, Computer modeling of plasma:  Past, present and future,  Phys. Plasmas {\bf 2} (1995) 2189-2199.

\bibitem{forslund1970}
D. W. Forslund and C. R. Shonk, Formation and structure of electrostatic collisionless shocks, Phys. Rev. lett. {\bf 25} (1970) 1699.

\bibitem{forslund1984} D. W. Forslund, K. B. Quest, J. U. Brackbill, K. Lee, Collisionless dissipation in quasi-perpendicular shocks, J. Geophys. Res. {\bf 89} (1984) 2142-2150.

\bibitem{lembege1989}
B. Lembege and J. M. Dawson, Formation of double layers within and oblique collisionless shock, Phys. Rev. Lett. {\bf 62} (1989) 2683-2686.

\bibitem{lembege2004}
B. Lembege, J. Giacolone, M. Scholer, T. Hada, Selected problems in collisionless shock physics,
Space Sci. Rev. {\bf 110} (2004) 161-236.

\bibitem{estabrook1983}
K. Estabrook and W. L. Kruer, Theory and simulation of one-dimensional Raman backward and forward scattering, Phys. Fluids {\bf 26} (1983) 1892.


\bibitem{forslund1975}
D. W. Forslund, J. M. Kindel, K. Lee, E. L. Lindman, R. L. Morse, Thoery and simulation of resonant absorption in a hot plasma,
Phys. Rev. A {\bf 11} (1975) 679.

\bibitem{forslund1982}
D. W. Forslund, J. U. Brackbill, Magnetic-field-induced surface transport on laser-irradiated foils,
Phys. Rev. Lett. {\bf 48} (1982) 1614.

\bibitem{wilks1992}
S. C. Wilks, W. L. Kruer, M. Tabak, A. B. Langdon,
Absorption of ultra-intense laser pulses,
Phys. Rev. Lett. {\bf 69} (1992) 1383.

\bibitem{pritchett2001}
P.L.Pritchett, Geospace Environment Modeling magnetic reconnection challenge:  Simulation with a full-particle electromagnetic code, J. Geophys. Res. {\bf 106} (2001) 3783-3798.



\bibitem{ricci2004} P. Ricci,  J. U. Brackbill, W. Daughton, G. Lapenta,  Influence of the lower hybrid drift instability on the onset of magnetic reconnection, Phys. Plasmas {\bf 11} (2004) 4489.

\bibitem{joshi1984}
C. Joshi {\it et al.}, Ultrahigh gradient particle acceleration by intense laser-driven plasma density waves, Nature {\bf 311} (1984) 525.

\bibitem{mangles2004}
S. P. D. Mangles {\it et al.}, Monoenergetic beams of relativistic electrons from intense laser-plasma interactions, Nature {\bf 431} (2004) 535-538.


\bibitem{villasenor1992}
J. Villasenor and O. Buneman,
Rigorous charge conservation for local electromagnetic field solvers,
Comput. Phys. Commun. {\bf69}(1992) 306-316.

\bibitem{chen2011}
G. Chen, L. Chacon, D. C. Barnes, An energy- and charge-conserving implicit electrostatic particle-in-cell algorithm, J. Comput. Phys. {\bf 230} (2011) 7018-7036.

\bibitem{denavit1981}
J. Denavit, Time-filtering particle simulations with $\omega_{pe}>>1$, J. Comput. Phys. {\bf 42}, 337-366 (1981).

\bibitem{mason1981} Rodney J. Mason, Implicit moment particle simulation of plasmas, J. Comput. Phys. {\bf 41} (1981) 233-244.

\bibitem{brackbill1982} J. U. Brackbill, D. W. Forslund, An implicit method for electromagnetic plasma simulation in two dimensions, J. Comput. Phys. {\bf 46} (1982)271.

 \bibitem{langdon1983}
A. B. Langdon, B. I. Cohen and A. Friedman, Direct implicit large time-step particle simulation of plasmas, {\bf 51}(1983)107-138.

\bibitem{markidis2011}
S. Markidis and G. Lapenta, The energy conserving particle-in-cell method, {\bf 230} (2011) 7037-7052.

\bibitem{harlow1959}
F. H. Harlow, D. O. Dickman, D. E. Harris, R. E. Martin, Los Alamos National Laboratory Report No. LA-2301, 1959 (unpublished).

\bibitem{harlow1964}
F. H. Harlow, The particle-in-cell computing method in fluid dynamics, Meth. Comput. Phys. {\bf 3} (1964) 319-343.

\bibitem{schindler1973}
K. Schindler, D. Pfirsch and H. Wobig,  Stability of two-dimensional collision-free plasmas,  Plasma Phys. {\bf 15} (1973) 1165-1184.

\bibitem{langdon1973}
A. B. Langdon, 'Energy conserving' plasma simulation, J. Comput. Phys. {\bf 12} (1973) 247-268.

\bibitem{vu1992}
H. X. Vu and J. U. Brackbill, CELEST1D: an implicit, fully kinetic model for low-frequency, electromagnetic plasma simulation, Comput. Phys. Commun. {\bf 69} (1992) 253-276.

\bibitem{sokolov2013}
I. V. Sokolov, Alternating-order interpolation in a charge-conserving scheme for particle-in-cell simulations,
Comput. Phys. Commun. {\bf 184} (2013) 320-328.




\bibitem{chacon2013}
L. Chacon, G. Chen, D. C. Barnes, A charge- and energy-conserving implicit, electrostatic particle-in-cell algorithm on mapped computational meshes, J. Comput. Phys. {\bf 233} (2013) 1-9.

\bibitem{langdon1970}
A. B. Langdon, Effects of spatial grids in simulation plasmas, {\bf 6} (1970) 247-267.

\bibitem{lindman1970}
 E. L. Lindman, Dispersion relation for computer-simulated plasmas, J. Comput. Phys. {\bf 5} (1970) 13-22.

\bibitem{landau1975}
L. D. Landau and E. M. Lifshitz, 'The Classical Theory of Fields', Fourth Revised English Edition, Elsevier, 1975.


\bibitem{brackbill1980}
J. U. Brackbill, D. C. Barnes, The effect of nonzero $\nabla \cdot \B$ on the numerical solution of the magnetohydrodynamic equations, J. Comput. Phys. {\bf 35} (1980) 426-430.



\bibitem{harlow1963}
F. H. Harlow, The particle-in-cell method for the numerical solution of problems in fluid dynamics,  in Proceedings of the Symposium on Appl. Math. XV, Experimental Arithmetic, High Speed Computations and Mathematics, American Math. Soc., Providence, 1963.

\bibitem{amsden1966}
A. A. Amsden, Los Alamos National Laboratory Report N. LA-3466, 1966  (unpublished).

\bibitem{brackbill1988}  J. U. Brackbill, The ringing instability in particle-in-cell calcuations of low-speed flow, J. Comput. Phys. {\bf 75} (1988) 469.

\bibitem{brackbill1986b} J. U. Brackbill and H. M. Ruppel, FLIP: A method for adaptively zoned, particle-in-cell calulations of fluid flows in two dimensions, J. Comput. Phys. {\bf 65}(1986) 314. 

\bibitem{deboor1978}
C. de Boor, 'A Practical Guide to Splines', Springer-Verlag, New York (1978).

\bibitem{haugbolle2013}
T. Haugbolle, J. T. Frederiksen, and A. Nordlund, 'Photon-Plasma:  A modern high-order particle-in-cell code,   Phys. Plasmas {\bf 20} (2013) 062904. 

\bibitem{lewis1970}
H. R. Lewis, Application of Hamilton's principle to the numerical analysis of Vlasov plasmas, in Meths. in Computational Physics, {\bf 9} (1970) 307-338.


\bibitem{goldstein1959}
H. Goldstein, 'Classical Mechanics', Addison-Wesley, Reading, Ma (1959).

\bibitem{okuda1972}
Hideo Okuda, Nonphysical noises and instabilities in plasma simulation due to a spatial grid, J. Comput. Phys. {\bf 10} (1972) 475-486.


\bibitem{birdsall1980}
C. K. Birdsall and N. Maron, Plasma self-heating and saturation due to numerical instabilities, J. Comput. Phys. {\bf 36} (1980) 1-19.

\bibitem{hockney1971}
R. W. Hockney, Measurements of collision and heating times in a two-dimensional thermal computer plasma, J. Comput. Phys. {\bf 8}(1971) 19-44.







\bibitem{hirose1982}
A. Hirose, O. Ishihara, and A. B. Langdon, Nonlinear evolution of the Buneman instability. II. Ion dynamics {\bf 25} (1982) 610.

\bibitem{cohen1989} 
Bruce I. Cohen, A. Bruce Langdon, Dennis W. Hewett, and Richard J. Procassini, Performance and optimization of direct implicit particle simulation, J. Comput. Phys. {\bf 81} (1989), 151-168.



























 
































































 

































\end{thebibliography}
\end{document}